\documentclass[]{article}
\usepackage[margin=.92in]{geometry}
\usepackage[utf8]{inputenc}
\usepackage{amsmath,color}
\usepackage{amsfonts}
\usepackage{mathrsfs}
\usepackage{mathtools}
\usepackage{graphicx}
\usepackage{bm}
\usepackage{wrapfig}
\usepackage{caption}
\usepackage{multicol}
\graphicspath{ {./images/} }
\numberwithin{equation}{section}

\usepackage[
backend=biber,
style=phys
]{biblatex}

\DeclarePairedDelimiter\bra{\langle}{\rvert}
\DeclarePairedDelimiter\ket{\lvert}{\rangle}
\DeclarePairedDelimiterX\braket[2]{\langle}{\rangle}{#1 \delimsize\vert #2}
\DeclarePairedDelimiterX\Braket[3]{\langle}{\rangle}{#1 \delimsize\vert #2 \delimsize\vert #3}

\newcommand{\SSigma}{\raisebox{-.25\baselineskip}{\huge{\ensuremath\Sigma}}}

\title{The Effect of Duschinskii Rotations on Spin-Dependent Electron Transfer Dynamics}
\author{Suraj S. Chandran, Yanze Wu, Joseph E. Subotnik}

\begin{document}

\maketitle

\begin{abstract}
    We investigate spin-dependent electron transfer in the presence of a Duschinskii rotation.  In particular, we propagate dynamics for a two-level model system for which spin-orbit coupling introduces an interstate coupling of the form $e^{iWx}$, which is both position($x$)-dependent and complex-valued. We demonstrate that two-level systems coupled to Brownian oscillators with Duschinskii rotations (and thus entangled normal modes) can produce marked increases in transient spin polarization relative to two-level systems coupled to simple shifted harmonic oscillators. These conclusions should have significant relevance for modeling the effect of nuclear motion on  chiral induced spin selectivity.
\end{abstract}

\section{Introduction}
Electron transfer is a crucial phenomenon underlying many aspects of chemistry, highlighting the fundamental energy conversion between electronic and nuclear degrees of freedom. As such, there has been an enormous push within the physical chemistry and chemical physics communities over the last 60 years to explore a variety of aspects related to electron transfer, including barriers to solution-phase thermal electron transfer \cite{jortner:1999:advchemphys,barbara:1996:marcus}, transient  photo-induced electron transfer dynamics \cite{tretiak:2014:acr}, electron-hole pair generation at metal interfaces\cite{suhl:1975:prb_elfriction,hynes:1993:elfriction,tully:1995:electronic_friction,wodtke:2000:science}, and much more.

The basic model for such processes is the shifted harmonic oscillator (SHO) model\cite{garg:onuchic:1985,leggett:1987:rmp} whereby we imagine two coupled harmonic surfaces, identical in all respects except for one: the well minima are shifted in space and found at different energies. The relevant Hamiltonian can be written as follows\cite{nitzanbook}:
\begin{align}
    \hat H &= \hat H_{nuc}+\hat H_{el}+\hat V\label{SB_model}\\
    \hat H_{nuc}&=\frac{\hat{\textbf p}\cdot\hat{\textbf p}}{2}+\frac12\hat{\textbf q}^T\hat{\bm\Omega}^2\hat{\textbf q}\\
    \hat H_{el} &= \hat H_D\ket d\bra d+\hat H_A\ket a\bra a\\
    \hat H_D &= E_D + \bm c_D^T\hat{\textbf q}\\
    \hat H_A &= E_A + \bm c_A^T\hat{\textbf q}\\
    \hat V &=V\ket d\bra a + V^*\ket a \bra d\label{SB_model_end}
\end{align}
Here, $\hat H_{nuc}$ and $\hat H_{el}$ respectively represent nuclear and electronic Hamiltonians. In $\hat H_{nuc}$, $\hat{\textbf p}$ and $\hat{\textbf q}$ are normal-mode momentum and position vector operators, and $\hat{\bm\Omega}^2$ is the (diagonal) harmonic matrix. In $\hat H_{el}$, $\ket d$ and $\ket a$ respectively represent donor and acceptor electronic states. The two states are coupled together by $V$ and coupled to the nuclear coordinates via the coupling vectors $\bm c_D$ and $\bm c_A$.  Taken together, Eqs. \ref{SB_model}-\ref{SB_model_end} embody the SHO model, otherwise known as the spin-boson model.

The spin-boson model has captivated researchers for decades, owing to the rich dynamics that emerge from its apparently simple construction\cite{reichman:2012:hybrid}. One regime of particular interest is that where the coupling $V$ between electronic states is small: the nonadiabatic regime\cite{nitzanbook,beratan:1990:jcp}. In a typical treatment of the nonadiabatic spin-boson model, one takes the initial state to be equilibrated in the donor well and applies perturbation theory to find the first-order correction to the acceptor-state population. In other words, the nuclear density matrix at time 0 is taken to be of the form
\begin{align}
    \hat\rho(0) &= \frac1Ze^{-\beta(\hat H_{nuc}+\hat H_D)}\\
    Z &= \text{Tr}\left[e^{-\beta(\hat H_{nuc}+\hat H_D)}\right]\label{part_func}
\end{align}
In the high-temperature limit, the resulting rate of transfer to the acceptor well is given by the classic Marcus rate expression:
\begin{equation}
    k = \frac{2\pi\lvert V\rvert^2}{\sqrt{4\pi E_rk_BT}}\exp\left(-\frac{(\Delta G+E_r)^2}{4E_rk_BT}\right)
\end{equation}
Here, $E_r$ is the reorganization energy, which defines the energy required to distort the nuclear geometry from the donor well minimum to the acceptor well minimum without an electronic transition.

Of course, the nonadiabatic regime is only one limit of the model; the spin-boson model can be used to investigate many other  regimes as well. For example, one might ask how the dynamics are different if the interstate coupling is large, a regime which is termed the adiabatic limit\cite{marcus:1963}. Alternatively, one might interrogate the solvent-controlled regime\cite{zusman:1980:original, hynes:1986:zusman,straub:1987:berne_nonadiabatic}, where strong interactions between the system and bath lead to heavy damping of the nuclear motions that produce electronic transitions. As yet another example, the spin-boson model can be extended beyond the basic formulation above to a model where the interstate coupling is dependent on nuclear coordinates, breaking the so-called Condon approximation\cite{jang:2005:noncondon, stuchebrukhov:1997:noncondon}.

In recent years, the discovery of the chiral induced spin selectivity (CISS) effect has begun to motivate investigation into yet another fundamental aspect of electron transfer\cite{naaman:2011:science:ciss_dna}: the realm of spin. A host of experiments within recent decades have shown that electron currents through chiral organic molecules can be strongly spin polarized\cite{naaman:2011:science:ciss_dna,naaman:2020:jpcl:perspective}. In this context, new questions arise: how can condensed-phase electron transfer exhibit such dramatic spin selectivity given the disproportionately small magnitude of organic spin-orbit couplings? Might such polarization be tied to nuclear geometry and motion\cite{Fay2021,fransson:2020prb:vibrational,opennheim2021} and, if so, how large can the resulting spin polarization be? And finally, in the spirit of the last few paragraphs, can the spin-boson model be an effective theoretical setting for probing such phenomena?

If we indeed seek to use the spin-boson model as a template for addressing the questions above, we will need to make one crucial modification. Namely, in order to capture the effects of SOC, we will need to explore regimes where the Hamiltonian is not real-valued, but rather complex-valued (i.e. where the off-diagonal coupling elements in Eq. \ref{SB_model_end} have non-zero imaginary components). To understand why this is so, consider the simple case of a intramolecular crossing between two doublets. This is a four-state system where a basis can be written as $\{\ket{1,\uparrow},\ket{1,\downarrow},\ket{2,\uparrow},\ket{2,\downarrow}\}$; in other words, we consider states formed by a direct product between two electronic molecular orbitals and two spin states. In this basis, the Hamiltonian can be written as follows (with $\hat{\textbf p}$ and $\hat{\textbf q}$ respectively denoting nuclear momenta and positions):

\begin{align}
    \label{doublet_crossing}\hat H &= \frac{\hat{\textbf p}\cdot\hat{\textbf p}}2+\hat {\mathcal V}(\hat{\textbf q})+\begin{bmatrix}
    E_1&0&V&0\\
    0&E_1&0&V\\
    V&0&E_2&0\\
    0&V&0&E_2
    \end{bmatrix}+\hat{\textbf L}\cdot\hat{\textbf S}\\
    \hat{\textbf L}\cdot\hat{\textbf S}&=\begin{bmatrix}
    0&0&\Braket{1}{\hat{\textbf L}_z}{2} &\Braket{1}{\hat{\textbf L}_x-i\hat{\textbf L}_y}{2}\\
    0&0&\Braket{1}{\hat{\textbf L}_x+i\hat{\textbf L}_y}{2}&-\Braket{1}{\hat{\textbf L}_z}{2}\\
    \Braket{2}{\hat{\textbf L}_z}{1} &\Braket{2}{\hat{\textbf L}_x-i\hat{\textbf L}_y}{1}&0&0\\
    \Braket{2}{\hat{\textbf L}_x+i\hat{\textbf L}_y}{1}&-\Braket{2}{\hat{\textbf L}_z}{1} &0&0
    \end{bmatrix}
\end{align}

In general, $\hat{\textbf L}_x$, $\hat{\textbf L}_y$, and $\hat{\textbf L}_z$ are all functions of $\hat{\textbf q}$. Notice that the $\hat{\textbf L}$ operators are proportional to momentum, so the matrix elements $\Braket{1}{\hat{\textbf L}}{2}$ are purely imaginary. Therefore, the two nonzero blocks of $\hat{\textbf L}\cdot\hat{\textbf S}$ are each anti-Hermitian and there exists a $\hat{\textbf q}$-dependent spin basis $\{\Tilde\uparrow(\hat{\textbf q}),\Tilde\downarrow(\hat{\textbf q})\}$ in which the blocks are diagonal with imaginary eigenvalues. In this basis, the Hamiltonian reads exactly as in Eq. \ref{doublet_crossing}, except that the spin-orbit operator now takes the form

\begin{equation}
    \hat{\textbf L}\cdot\hat{\mathcal S}=\begin{bmatrix}
    0 & 0 & i\alpha(\hat{\textbf q}) & 0\\
    0 & 0 & 0 & -i\alpha(\hat{\textbf q})\\
    -i\alpha(\hat{\textbf q}) & 0 & 0 & 0\\
    0 & i\alpha(\hat{\textbf q}) & 0 & 0
    \end{bmatrix}
\end{equation}

Here, $\alpha$ is a real-valued function of $\hat{\textbf q}$. If we now make the assumption that the spin basis is \textit{independent} of $\hat{\textbf q}$ (i.e. the Hamiltonian is spin-conserving) and project the Hamiltonian onto the two spin-states, we obtain two Hamiltonians, one of which governs the evolution of $\ket{\Tilde{\uparrow}}$ states, and another which governs the evolution of $\ket{\Tilde{\downarrow}}$ states:

\begin{align}
    \hat H(\Tilde{\uparrow}) &= \Braket{\Tilde\uparrow}{\hat H}{\Tilde \uparrow} = \frac{\hat{\textbf p}\cdot\hat{\textbf p}}2+\hat {\mathcal V}(\hat{\textbf q})+\begin{bmatrix}
    E_1&V+i\alpha(\hat{\textbf q})\\
    V-i\alpha(\hat{\textbf q})&E_2\\
    \end{bmatrix}\label{up_H}\\
    \hat H(\Tilde{\downarrow}) &= \Braket{\Tilde\downarrow}{\hat H}{\Tilde \downarrow} = \frac{\hat{\textbf p}\cdot\hat{\textbf p}}2+\hat {\mathcal V}(\hat{\textbf q})+\begin{bmatrix}
    E_1&V-i\alpha(\hat{\textbf q})\\
    V+i\alpha(\hat{\textbf q})&E_2\\
    \end{bmatrix}=\hat H(\Tilde\uparrow)^*
\end{align}

In other words, for a spin-conserving process occurring in the presence of spin-orbit coupling, spin-up and spin-down electrons evolve according to \textit{different} complex-valued Hamiltonians: if the former evolve according to $\hat H$, the latter evolve according to $\hat H^*$.

Now what are the implications of propagating a complex-valued Hamiltonian? At first glance, one might expect equivalent dynamics to arise regardless of whether one propagates using $\hat H$ or $\hat H^*$. After all, the eigenvalues of both operators are exactly the same. Indeed, if we consider equilibrium Fermi Golden Rule rates from donor to acceptor for a general vibronic system, it is clear that there can be no difference in $\hat H$ vs. $\hat H^*$ dynamics:
\begin{align}\label{fgr}
    k_{d\rightarrow a}(\hat H) &= \sum_{\textbf v,\textbf v'}P_{\textbf v}\lvert\Braket{a,\textbf v'}{\hat V}{d,\textbf v}\rvert^2\delta(E_{a,\textbf v'}-E_{d,\textbf v})\\
    k_{d\rightarrow a}(\hat H^*) &= \sum_{\textbf v,\textbf v'}P_{\textbf v}\lvert\Braket{a,\textbf v'}{\hat V^*}{d,\textbf v}\rvert^2\delta(E_{a,\textbf v'}-E_{d,\textbf v})=k_{d\rightarrow a}(\hat H)
\end{align}
Here, $\textbf v'$ and $\textbf v$ respectively index the acceptor and donor vibrational eigenstates, and $P_{\textbf v}=\frac{1}Z\exp(-\beta E_{d,\textbf v})$ with $Z$ defined as in \ref{part_func}.

However, upon reflection, one should realize that these conclusions hold only to the lowest order of perturbation theory; furthermore, they rely on the initial conditions being those of equilibrium\cite{wu:2020:jpca:spin}. Beyond these limits, a subtle difference in the properties of $\hat H$ and $\hat H^*$ becomes critically important; namely the fact that the respective eigenvectors differ by complex conjugation. From a semiclassical (rather than purely quantum) perspective, this subtle difference can lead to very different so-called ``Berry forces'' for  $\hat H$ vs $\hat H^*$ trajectories. As a review for the reader, in the adiabatic limit, semiclassical trajectories on adiabat $n$ experience a Berry force given by the following equation\cite{berry:1993:royal:half_classical, subotnik:2019:jcp_berry_qcle}, :
\begin{equation}\label{berry}
    \textbf{F}_{B}^{(n)}=i\hbar\dot{\textbf R}\times(\nabla_{\textbf R}\times\textbf D_{nn})
\end{equation}
Here, $\nabla_{\textbf R}$ is the gradient operator with respect to nuclear coordinates and $\textbf D_{nn}$ is the derivative coupling from adiabatic surface $n$ to itself.  
Note that the Berry force vector is orthogonal to the momentum of the nuclei ($\textbf F_B\cdot\dot{\textbf R}=0$) and can be understood semiclassically as an effective magnetic field that acts on nuclei near crossing points.
Note also that, according to Eq. \ref{berry}, it follows that the adiabatic Berry forces for $\hat H$ are equal and opposite to those for $\hat H^*$. which should appear reasonable insofar as flipping the spin of the relevant electron(s) effectively reverses all magnetic fields in the system.
 The take home point is that, if we run nuclera trajectories for systems with spin up and spin down electrons, we {\em will} find different dynamics; the magnitude of these differences will necessarily depend on many factors, including the topography of the potential energy surfaces as well as the size of the Berry force.  
 
 At this point, a few words are appropriate regarding the size and meaning of the Berry force. In general,  one should apply Eq. \ref{berry}  to a trajectory only in the adiabatic limit; Eq. \ref{berry} is derivable only when 
 there is a large energy gap between adiabatic electronic states.
That being said, the astute reader will notice that in the nonadiabatic regime, the derivative couplings (and therefore Eq. \ref{berry}) happen to be very large near electronic crossing points; one is then inevitably compelled to ask whether or not such large matrix elements might have a significant effect on entangled nuclear-electronic-spin quantum dynamics.  Might such Berry forces (arising from Hamiltonians with small SOC matrix elements) have large effects on dynamics and  therefore (at least partly) mediate the CISS effect? Unfortunately, at present, very little is known about the meaning of the forces in Eq. \ref{berry} in the nonadiabatic regime. To that end, our research group has recently begun constructing novel, nonadiabatic surface hopping approaches with the eventual goal of treating {\em ab initio} nuclear dynamics in the presence of a set of degenerate  electronic spin states.  

Looking forward, until such a semiclassical {\em ab initio} algorithm is available\cite{bian:wu:2022:pssh,roel:reichman:2018:jcp}, we believe the best approach to evaluating the size and impact of Berry forces is to develop an analytic model where intuition can be easily gleaned and relied upon. 
To that end, in Ref. \cite{chandran:spinrates:2022}, we  studied a spin-boson model with an Exponential Spin-Orbit Coupling (ESOC).  For such a problem, the Hamiltonian is defined exactly as in Eq. \ref{SB_model} with the only change being that the interstate coupling operator $\hat V$ takes the form complex-valued non-Condon form:
\begin{equation}\label{ESOC}
    \hat V = Ve^{i\textbf W^T\hat{\textbf q}}\ket d\bra a+V^*e^{-i\textbf W^T\hat{\textbf q}}\ket a\bra d
\end{equation}
Here, $\textbf W$ can be understood as encoding the gradient of the spin-orbit coupling near a diabatic crossing point. To illustrate this, we expand the coupling operator in Eq. \ref{ESOC} to first order in $\hat{\textbf q}$ around a crossing point $\textbf q_0$:
\begin{equation}\label{coupling_linear}
    Ve^{i\textbf W^T\hat{\textbf q}}\simeq Ve^{i\textbf W^T\textbf q_0}(1+i\textbf W^T(\hat{\textbf q}-\textbf q_0))
\end{equation}
Eq. \ref{coupling_linear} can be directly compared with 
$V+i\alpha(\hat{\textbf q})$ from Eq. \ref{up_H}, justifying the use of this Hamiltonian.

Using the Hamiltonian defined by Eqs. \ref{SB_model} and \ref{ESOC}, we demonstrated in Ref. \cite{chandran:spinrates:2022} that spin polarization could indeed emerge dynamically as a consequence of \textit{nonequilibrium} initial conditions. We note, however, that these conclusions required broken mirror symmetry; if a system is mirror symmetric, then the equal and opposite Berry forces produced in the respective $\hat H$ and $\hat H^*$ systems produce equivalent, albeit mirrored, nuclear dynamics (see below). These results would appear consistent with the fact that experimentally, for the CISS effect, one must run current through a chiral molecule in order to achive spin separation in general (and of course, chiral molecules, by definition, have no mirror symmetry).

With this background in mind, the goal of the present article is to gain intuition about the about the extent of the expected transient spin-polarized electron transfer that will occur in the absence of mirror symmetry when we go beyond the constraints of the SHO model. In particular, we will study if/how the presence of intertwined vibrational modes (as modeled with a Duchinskii rotation) affects spin selectivity. To achieve this goal, we will investigate the ESOC model in the context of an intuitive model electron-transfer system where two coupled primary modes are each interacting with a large dissipative thermal bath \cite{leggett:1983:caldeira}. Working within this framework, we will answer two questions.
First, how does the direction of the intramolecular spin-orbit coupling vector influence the spin-polarization of the final state? Here, the partitioning between primary and bath modes proposed above is especially useful, since the primary modes can be thought of as defining a "molecule of interest" within our very large system. In this context, the concepts of intramolecular and intermolecular interactions are well-defined; any coupling to the primary modes is an intramolecular interaction, whereas any coupling to the bath modes is an intermolecular interaction.
Second, can systems in which normal mode frequencies change upon electronic transition produce much stronger spin polarization than simple Shifted Harmonic Oscillator (SHO) model systems, and if so, when are these effects maximal? Answering these questions will inevitably give us a great deal of insight into the coupled nature of nuclear-electronic-spin dynamics.

An outline of this article is as follows. First, in Sec. \ref{Theory}, we present the relevant theoretical background needed to construct the aforementioned model system and to model its dynamics using a Duschinskii rotation formalism \cite{pollak:2004:duschinskii_cooling, duschinskii:1937}. Next, in Sec. \ref{Results} we present results on how changing the geometry of a two-state harmonic system via Duschinskii rotations affects spin polarization. Finally, in Sec. \ref{Discussion} we discuss these results in the context of mirror symmetry and chirality.

\section{Theory}\label{Theory}

\subsection{The Definition of Duschinskii Rotations for a General Two-State Quadratic Hamiltonian}\label{dusc_mot}

Consider a general quadratic Hamiltonian with a potential of the following form:
\begin{align}\label{quad_H}
    \hat{\mathcal V} = \begin{bmatrix}
        \frac12\hat{\textbf q}^T\hat{\bm\Omega}_g^2\hat{\textbf q}+\bm\lambda_g^T\hat{\textbf q}+E_g & \hat V(\hat{\textbf q})\\
        \hat V(\hat{\textbf q})^\dagger & \frac12\hat{\textbf q}^T\hat{\bm{\Omega}}_e^2\hat{\textbf q}+\bm\lambda_e^T\hat{\textbf q}+E_e
    \end{bmatrix}
\end{align}
Here, $\hat{\textbf q}$ is a vector operator consisting of the position operators for each mode in the system and $g$, $e$ respectively denote the ground and excited states. The Hamiltonian defined by Eq. \ref{quad_H} is a generalized spin-boson model where the harmonic surfaces of the ground and excited state diabats are not required to be identical. The two diabatic harmonic matrices $\hat{\bm\Omega}_g^2$ and $\hat{\bm\Omega}_e^2$ are symmetric and can be diagonalized respectively by orthogonal matrices $\textbf S_g$ and $\textbf S_e$: 
\begin{align}
    \hat{\bm \Omega}_g&=\textbf S_g\hat{\bm\Delta}_g\textbf S_g^T\\
    \hat{\bm \Omega}_e&=\textbf S_e\hat{\bm\Delta}_e\textbf S_e^T    
\end{align}
Therefore, we can write the diabatic Hamiltonians as follows:
\begin{align}
    \frac12\hat{\textbf q}^T\hat{\bm\Omega}_g^2\hat{\textbf q}+{\bm\lambda}_g^T\hat{\textbf q}&=\frac12(\hat{\textbf q}^T+{\bm\lambda}_g^T\hat{\bm\Omega}_g^{-2})\textbf S_g\hat{\bm\Delta}_g^2\textbf S_g^T(\hat{\textbf q}+\hat{\bm\Omega}_g^{-2}{\bm\lambda}_g)-\frac12{\bm\lambda}_g^T\hat{\bm\Omega}_g^{-2}{\bm\lambda}_g+E_g\\
    \frac12\hat{\textbf q}^T\hat{\bm\Omega}_e^2\hat{\textbf q}+{\bm\lambda}_e^T\hat{\textbf q}&=\frac12(\hat{\textbf q}^T+{\bm\lambda}_e^T\hat{\bm\Omega}_e^{-2})\textbf S_e\hat{\bm\Delta}_e^2\textbf S_e^T(\hat{\textbf q}+\hat{\bm\Omega}_e^{-2}{\bm\lambda}_e)-\frac12{\bm\lambda}_e^T\hat{\bm\Omega}_e^{-2}{\bm\lambda}_e+E_e
\end{align}

The matrices $\hat{\bm\Delta}$ are diagonal operators, and they encode the normal mode frequencies for the upper and lower diabats. Now, let us substitute $\hat{\textbf q}'=\textbf S_g^T(\hat{\textbf q}+\hat{\bm\Omega}_g^{-2}{\bm\lambda}_g)$ and $\hat{\textbf q}''=\textbf S_e^T(\hat{\textbf q}+\hat{\bm\Omega}_e^{-2}{\bm\lambda}_e)$. In this new basis, the two blocks become:
\begin{align}
    &\frac12\hat{\textbf q}'^T\hat{\bm\Delta}_g^2\hat{\textbf q}'-\frac12{\bm\lambda}_g^T\hat{\bm\Omega}_g^{-2}{\bm\lambda}_g\label{q'}+E_g\\
    &\frac12\hat{\textbf q}''^T\hat{\bm\Delta}_e^2\hat{\textbf q}''-\frac12{\bm\lambda}_e^T\hat{\bm\Omega}_e^{-2}{\bm\lambda}_e+E_e \label{q''}
\end{align}

Also note that the position operators $\textbf q'$ and $\textbf q''$ are related by the following transformation:
\begin{equation}
    \hat{\textbf q}'=\textbf S_g^T\textbf S_e\hat{\textbf q}''+\textbf S_g^T(\hat{\bm\Omega}_g^{-2}{\bm\lambda}_g-\hat{\bm\Omega}_e^{-2}{\bm\lambda}_e)\label{q_transform}
\end{equation}

The ground state normal mode vector $\hat{\textbf q}'$ is of the form $\textbf{S}\hat{\textbf q}''+\textbf d$. The matrix $\textbf S=\textbf S_g^T\textbf S_e$ is known as a Duschinskii rotation matrix \cite{pollak:2004:duschinskii_cooling} and describes how the normal modes of one diabat can be expressed as linear combinations of normal modes of the other diabat. In a shifted harmonic oscillator system, the Duschinskii matrix would simply be the identity since both diabats would have identical normal modes. The vector $\textbf d =\textbf S_g^T(\hat{\bm\Omega}_g^{-2}\bm\lambda_g-\hat{\bm\Omega}_e^{-2}\bm\lambda_e)$ in Eq. \ref{q_transform} is the shift vector, and describes the displacement between the well minima of the two diabats. Note that, under the transformation in Eqs. \ref{q'} and \ref{q''}, $\hat V(\hat{\textbf q})$ is transformed to $\hat V(\textbf S_e\hat{\textbf q}''-\hat{\bm\Omega}_e^{-2}\bm\lambda_e)$; in particular, if $\hat V=e^{i\textbf W^T\hat{\textbf q}}$, $\hat V$ is mapped to $e^{i\textbf W^T \textbf S_e\hat{\textbf q}''}$ (up to a constant complex phase). Finally, note also that energy difference between the ground and excited state minima ($\Delta G$) can be written as $E_g-E_e+\frac12\left(\bm\lambda_e^T\hat{\bm\Omega}_e^{-2}\bm\lambda_e-\bm\lambda_g^T\hat{\bm\Omega}_g^{-2}\bm\lambda_g\right)$.

We will use the parametrization in Eqs. \ref{q'} and \ref{q''} to model the dynamics of molecular electron transfer systems for which the vibrational modes may or may not change over the course of the event.

\subsection{Dynamics of Systems with Duschinskii Rotations}\label{Dusc_dynamics}

We have applied the Fermi Golden Rule formalism to the Hamitonian above. The lengthy calculation \cite{pollak:2004:duschinskii_cooling} is provided in the appendix. We focus on a Hamiltonian of the following form:
\begin{align}
    \hat{\mathcal H} &= \begin{bmatrix}
    \hat H_B^{(g)} & Ve^{i\textbf W^T\hat{\textbf x}}\\
    V^*e^{-i\textbf W^T \hat{\textbf x}} & \hat H_B^{(e)}
    \end{bmatrix}\label{duschinskii_H}\\
    \hat H_B^{(g)} &= \frac{\hat{\textbf p}\cdot\hat{\textbf p}}{2}+\frac12\left(\textbf S\hat{\textbf x}+\textbf d\right)^T\hat{\bm\Omega}_g^2\left(\textbf S\hat{\textbf x}+\textbf d\right) + \Delta G\label{H_B^g}\\
    \hat H_B^{(e)} &= \frac{\hat{\textbf p}\cdot\hat{\textbf p}}{2}+\frac12\hat{\textbf x}^T\hat{\bm\Omega}_e^2\hat{\textbf x}
\end{align}
here, $\textbf S$ is the Duschinskii rotation matrix, $\textbf d$ is the displacement vector between the well minima, and the harmonic matrices $\hat{\bm\Omega}_e$ and $\hat{\bm \Omega}_g$ are assumed to be diagonal. Before proceeding further, it will be helpful to define several matrix functions which will appear below:

\begin{multicols}{2}
    \begin{equation}
        \bm a_e(t) = \hat{\bm\Omega}_e[\sin(\hat{\bm\Omega}_et)]^{-1}
    \end{equation}
    \begin{equation}
        \bm a_g(t) = \hat{\bm\Omega}_g[\sin(\hat{\bm\Omega}_gt)]^{-1}
    \end{equation}
    \begin{equation}
        \bm b_e(t) = \hat{\bm\Omega}_e[\tan(\hat{\bm\Omega}_et)]^{-1}
    \end{equation}
    \begin{equation}
        \bm b_g(t) = \hat{\bm\Omega}_g[\tan(\hat{\bm\Omega}_gt)]^{-1}
    \end{equation}
    \begin{equation}
        \textbf A = \bm a_e(-\tau-i\beta)+\textbf S^T \bm a_g(\tau)\textbf S
    \end{equation}
    \begin{equation}
        \textbf B = \bm b_e(-\tau-i\beta)+\textbf S^T\bm b_g(\tau)\textbf S
    \end{equation}
    \begin{equation}
        \textbf E = \bm b_e(-\tau-i\beta)-\bm a_e(-\tau-i\beta)
    \end{equation}
    \begin{equation}
        \textbf G = \bm b_g(\tau)-\bm a_g(\tau)
    \end{equation}
    \begin{equation}
        \textbf{G}_\beta = \bm b_g(-i\beta)-\bm a_g(-i\beta)
    \end{equation}
\end{multicols}

\subsubsection{Equilibrium Dynamics}

Let us begin by initializing the system on the excited state with the equilibrium density matrix
\begin{align}
    \hat \rho(0) &= \frac 1Ze^{-\beta \hat H_B^{(e)}}\ket e\bra e\\
    Z &= \text{Tr}_B\left[e^{-\beta\hat H_B^{(e)}}\right]
\end{align}
The resulting correlation function that defines the rate of electron transfer to the ground state of the system is given by (see App. \ref{app_eq}):
\begin{align}
    C(\tau)&=\langle \hat V_{eg}(\tau)\hat V_{ge}\rangle\\
    &=\frac{1}{Z}\text{Tr}_B\left[e^{-i\hat H_B^{(e)}(-\tau-i\beta)}e^{-i\textbf W^T\hat{\textbf x}}e^{-i\hat H_B^{(g)}\tau}e^{i\textbf W^T \hat{\textbf x}}\right]\\
    &=\frac{1}{Z}\sqrt{\frac{\det(\bm a_e(-\tau-i\beta))\det(\bm a_g(\tau))}{\det(\textbf B)\det(\textbf B-\textbf{AB}^{-1}\textbf A)}}\exp\left(i\left[\textbf d^{T}\textbf{GS}(\textbf B-\textbf A)^{-1}\textbf{ES}^{T}\textbf d-\textbf W^{T}(\textbf B+\textbf A)^{-1}\textbf W\right]\right)\label{eq_corr}
\end{align}

Notably, the term in the exponential has a bilinear dependence on $\textbf W$, meaning that the sign of $\textbf W$ has no bearing on the equilibrium dynamics.  This suggests that if there were to be a dynamical dependence on the sign of $\textbf W$ for equilibrium dynamics, such an effect should be quite small because it could reveal itself only at higher orders of perturbation theory. Following standard perturbation theory, the ground state population at time $t$ is given by:
\begin{equation}
    P_g(t)\simeq 2\lvert V\rvert^2\cdot\text{Re}\left[\int_0^tdt'\,\int_0^{t'}d\tau\,e^{-i\Delta G\tau}C(\tau)\right]
\end{equation}
where $\Delta G$ is defined in Eq. \ref{H_B^g}.

\subsubsection{Nonequilibrium Dynamics}

Our primary motivation in developing a nonequilibrium formalism for Duschinskii systems is to understand the consequences of Duschinskii rotations with regards to photoexcitation dynamics. To that end, let us next focus on the case where the system begins in the excited state with the density matrix
\begin{align}
    \hat \rho_{neq}(0) &= \frac 1Ze^{-\beta \hat H_B^{(g)}}\ket e\bra e\\
    Z &= \text{Tr}\left[e^{-\beta\hat H_B^{(g)}}\right]
\end{align}
Notice here that the thermal distribution is based on the \textit{ground} state Hamiltonian; the intuition is that, prior to a fast photoexcitation, we expect the nuclear modes of the system to be equilibrated in the ground electronic state. When the fast photoexcitation occurs, the comparatively slow timescale of the nuclear vibrations implies that the nuclei will not have time to equilibrate according to the new electronic state; instead, dynamics will launch from a state where the nuclei are still in the ground state geometry.

In the nonequilibrium case, the correlation function is significantly more complicated and cannot be reduced to a function of a single time variable but rather depends on two times:
\begin{align}
    C(t',t'') &=\text{Tr}_B\left[\hat\rho_{neq}(0)\hat V_{eg}(t')\hat V_{ge}(t'')\right]\\
    &= \text{Tr}_B\left[e^{-\beta\hat H_B^{(g)}}e^{i\hat H_B^{(e)}t'}e^{-i\textbf W^T\hat{\textbf x}}e^{-i\hat H_B^{(g)}(t'-t'')}e^{i\textbf W^T \hat{\textbf x}}e^{-i\hat H_B^{(e)}t''}\right]\\
    &=\frac 1Z\sqrt{\frac{\det\left[\bm a_g(-i\beta)\bm a_e(-t')\bm a_g(t'-t'')\bm a_e(t'')\right]}{\det\Sigma}}\nonumber\\
    &\qquad\times\exp\left(i\textbf d^{T}\left[\textbf{G}_\beta+\textbf G\right]\textbf d-\frac12i\begin{bmatrix}
    \textbf S^T\textbf{G}_\beta\textbf d\\
    \textbf S^T\textbf{G}_\beta\textbf d\\
    \textbf S^{T}\textbf G\textbf d-\textbf W\\
    \textbf S^{T}\textbf G\textbf d+\textbf W
    \end{bmatrix}^{T}\SSigma^{-1}\begin{bmatrix}
    \textbf {S}^T\textbf{G}_\beta\textbf d\\
    \textbf {S}^T\textbf{G}_\beta\textbf d\\
    \textbf S^T\textbf G\textbf d-\textbf W\\
    \textbf S^T\textbf G\textbf d+\textbf W
    \end{bmatrix}\right)\label{neq_corr}
\end{align}
\begin{equation}
    \small{\SSigma = \begin{bmatrix}
    \textbf S^T\bm b_g(-i\beta)\textbf S+\bm b_e(t'')&-\bm a_g(-i\beta)&0&-\bm a_e(t'')\\
    -\bm a_g(-i\beta)&\textbf S^T\bm b_g(-i\beta)\textbf S+\bm b_e(-t')&-\bm a_e(-t')&0\\
    0&-\bm a_e(-t')&\bm b_e(-t')+\textbf S^{T}\bm b_g(t'-t'')\textbf S&-\textbf S^{T}\bm a_g(t'-t'')\textbf S\\
    -\bm a_e(t'')&0&-\textbf S^{T}\bm a_g(t'-t'')\textbf S &\bm b_e(t'')+\textbf S^{T}\bm b_g(t'-t'')\textbf S
    \end{bmatrix}}
\end{equation}

Again, the correlation function can be used to write down an expression for the approximate ground-state population at time $t$:
\begin{equation}
    P_g(t)\simeq \lvert V\rvert^2\int_0^tdt'\,\int_0^{t}dt''\,e^{-i\Delta G(t''-t')}C(t',t'')
\end{equation}

\subsection{An Intuitive Model with Two Intertwined Modes and Nuclear Friction }\label{lang_map}

The above formulation is rather abstract, and it is therefore difficult to properly interpret the obtained equations and their consequences. Therefore, it will be helpful to define an intuitive test model consisting of two intertwined nuclear modes (primary modes) each coupled to a large thermal bath of secondary modes. The exact coupling to the  bath of secondary modes will be defined so as to produce Langevin dynamics for each primary nuclear mode. It is easy to see how such a model maps onto a real condensed-phase chemical system; one can imagine that the primary modes represent a molecule of interest undergoing electron transfer dynamics while interacting with a solvent environment, which is described by the secondary modes. Mathematically, we describe this system using the following Hamiltonian:

\begin{align}\label{langevin_H}
    \hat{\mathcal H} = &\frac{\hat{\textbf p}\cdot\hat{\textbf p}}{2}+\hat H_B^{(x)}+\hat H_B^{(y)}+\\
    &\begin{bmatrix}
        \frac12\omega_2^2\left(\hat x-d\cos\theta\right)^2+\frac12\omega_1^2\left(\hat y-d\sin\theta\right)^2+E_g& Ve^{iW(\hat x\cos\eta+\hat y\sin\eta)}\\
        V^*e^{-iW(\hat x\cos\eta+\hat y\sin\eta)}& \frac12\omega_1^2(\hat x\cos\phi+\hat y\sin\phi)^2+\frac12\omega_2^2(\hat y\cos\phi-\hat x\sin\phi)^2+E_e\nonumber
    \end{bmatrix}
\end{align}
\begin{align*}
    \hat H_B^{(x)} = \sum_\alpha \frac12\omega_\alpha^2\left(\hat q_\alpha+\frac{c_\alpha\hat x}{\omega_\alpha^2}\right)^2 \qquad \hat H_B^{(y)} = \sum_\beta \frac12\omega_\beta^2\left(\hat q_\beta+\frac{c_\beta\hat y}{\omega_\beta^2}\right)^2
\end{align*}

Here, $\hat x$ and $\hat y$ are the position operators for the primary modes, and are respectively coupled to the thermal baths given by$\hat H_B^{(x)}$ and $\hat H_B^{(y)}$ (the exact couplings $c_\alpha$ and $c_\beta$ will be defined below). These baths are responsible for producing random fluctuations as well as an apparent friction in the motion of the primary modes.

\begin{figure}
    \centering
    \captionsetup{width=.83\linewidth}
    \includegraphics[scale=0.7]{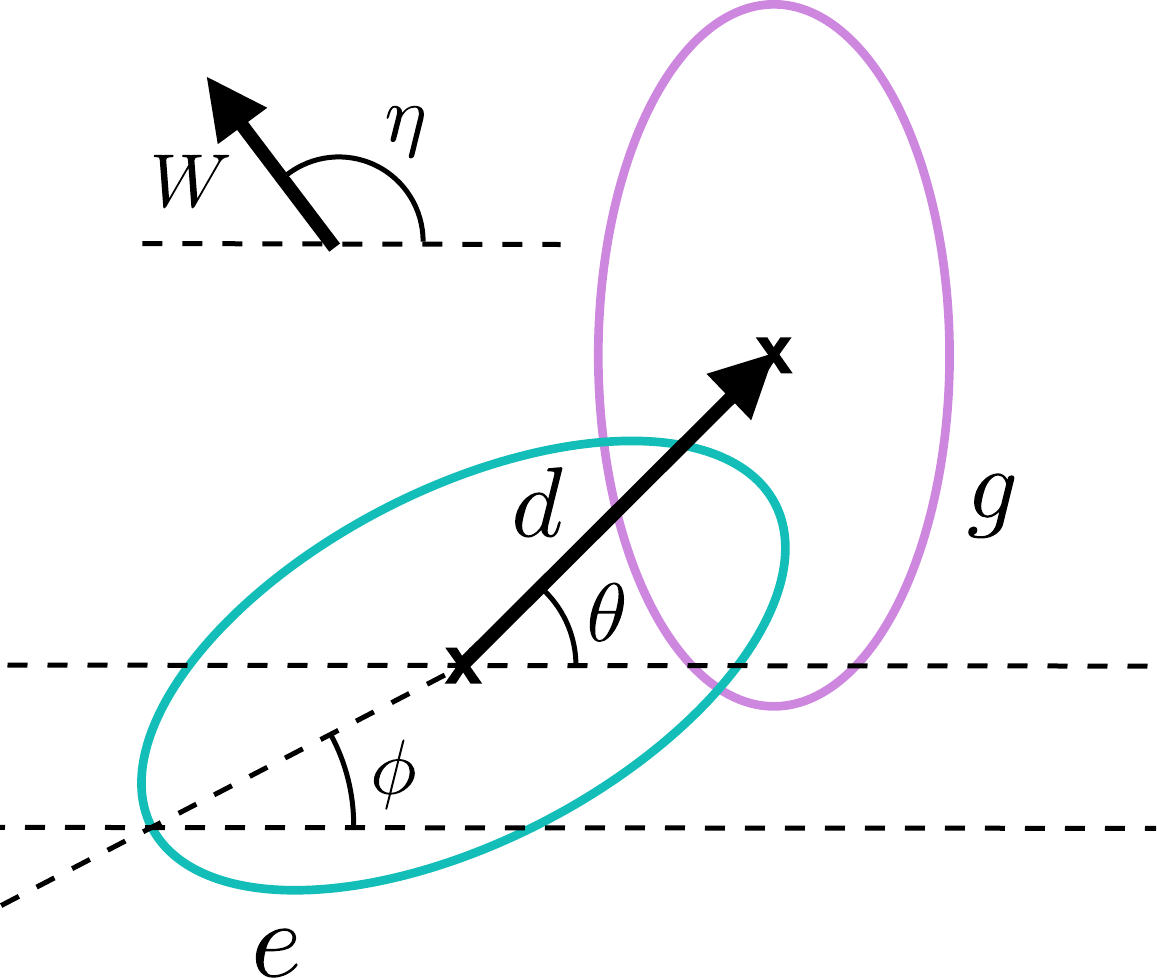}
    \caption{Shown is an example of a system that may be defined by the Hamiltonian in Eq. \ref{langevin_H}. The parameters $d$, $W$, $\theta$, $\eta$, and $\phi$ are shown in context. $d$ and $\theta$ together define the shift between the minima of the primary mode wells. Similarly, $W$ and $\eta$ together define the spin-orbit coupling vector within the primary system. Finally, $\phi$ defines the rotation angle on the excited state well, as measured from the major axis to the horizontal $x-$axis.}
    \label{wells}
\end{figure}

There are three main parameters defining the geometry of the primary-mode system (i.e. the molecule of interest): $\phi$, $\theta$, and $\eta$. The excited-state well (the lower right block of Eq. \ref{langevin_H}) is centered on the origin and has a rotation angle defined by $\phi$ relative to the $x-$axis. The minimum of the ground-state well (the upper left block of Eq. \ref{langevin_H}) is shifted away from that of the excited-state minimum along an axis defined by the angle $\theta$ relative to the $x-$axis. Together, $\phi$ and $\eta$ define the spatial geometry of the primary-mode system; the well geometries are plotted for a few different values of $\phi$ and $\theta$ in Fig. \ref{wells}. Finally, $\eta$ describes the direction of the spin-orbit coupling vector. These three parameters, together with the magnitudes of the shift and spin-orbit coupling vectors ($d$ and $W$) are enough to specify the vector parameters $\textbf d$, $\textbf S$, and $\textbf W$ (respectively the shift vector, Duschinskii matrix, and spin-orbit coupling vector) from Section \ref{Dusc_dynamics}. 

We have imposed several important and deliberate constraints in the construction of Eq. \ref{langevin_H}. First, note that the interstate coupling operator depends purely on $\hat x$ and $\hat y$; it is wholly independent from the secondary bath modes. This constraint reflects the physical notion that the environment of a molecule has no influence on the spin-orbit coupling within the molecule. Such an assumption, while likely not universally true in real systems, is a reasonable approximation we have made for the sake of gaining basic intuition about the problem at hand.

\begin{figure}
    \centering
    \includegraphics[scale=0.5]{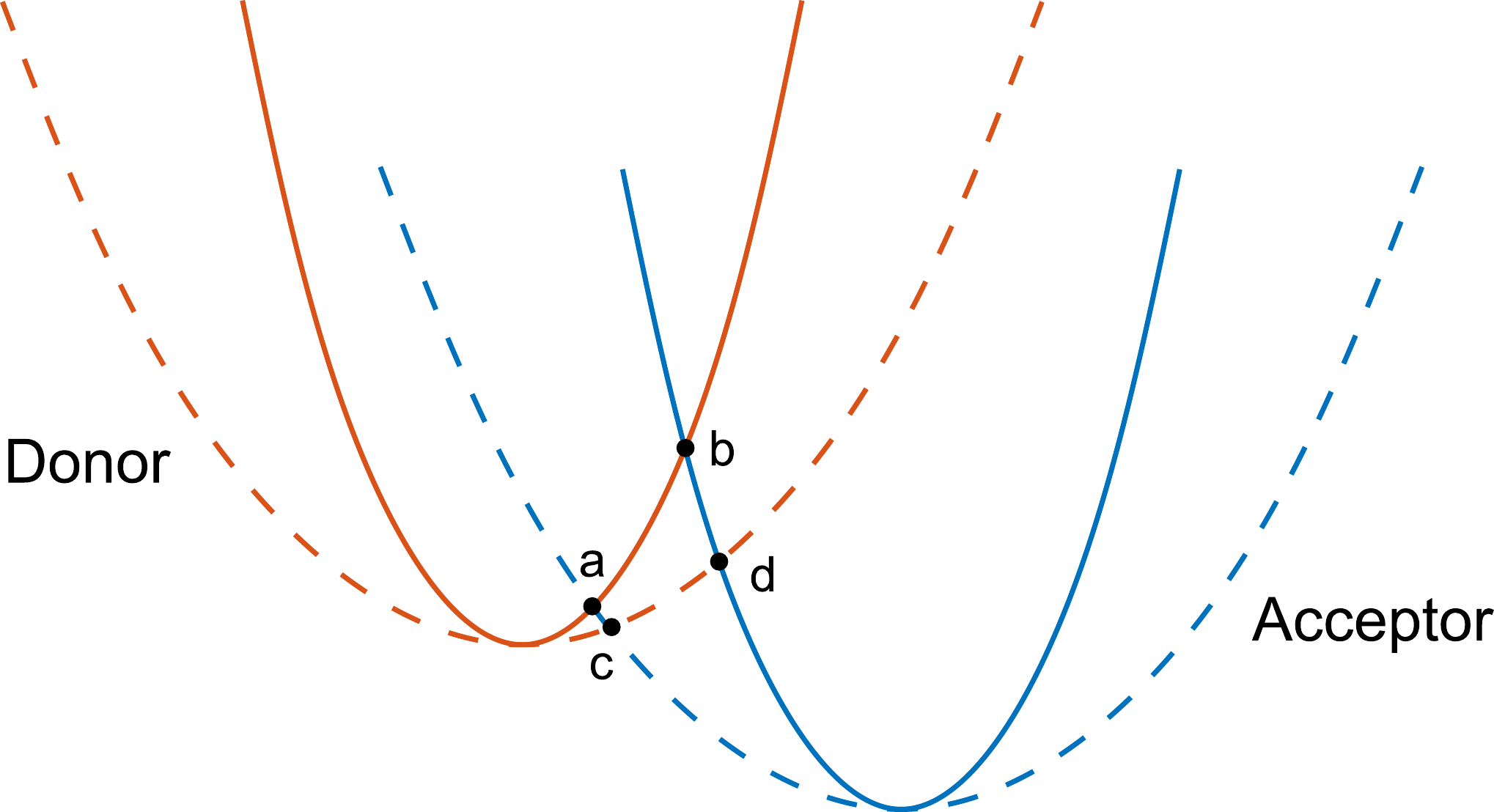}
    \caption{Pictured here is a simple one-dimensional two-state vibronic system. Notice that the donor$\rightarrow$acceptor activation barrier is more sensitive to the acceptor state frequency than it is to the donor state frequency. This is illustrated by the observation that the changes in energy from $\bm a$ to $\bm c$ and $\bm b$ to $\bm d$ are not as significant as those from $\bm a$ to $\bm b$ and $\bm c$ to $\bm d$. By analogy, we expect that the activation barrier for a much larger multidimensional will be more sensitive to the geometry of the acceptor state well than that of the donor state well. Therefore, to ensure a similar timescale for electron transfer across a range of Hamiltonians, we will use the parameter $\phi$ to rotate the excited state well rather than the ground state well.}
    \label{rotation_Er}
\end{figure}

Second, notice that we have deliberately chosen to use the coordinate $\phi$ (which represents rotations of the excited state well) to characterize each Hamiltonian.  In general, note that rotating the excited state does not lead to drastically different ranges of electron transfer rates; rotating the ground state, however, would lead to drastically different ranges of electron transfer rates. Therefore, in order to model dynamics over a range of different geometries -- but with a reasonable range of time scales -- for a fixed $\theta$, we will simply vary $\phi$ below and record the resulting dynamics. More generally, even for a set of displaced multidimensional harmonic oscillators with different frequencies, a reasonable energy scale can be calculated the driving force that renders the electron transfer barrierless; notice that this driving force depends only on the frequencies of the acceptor molecule (and not on the frequencies of the donor molecule). Thus, changing $\phi$ (while keeping $\theta$ fixed) is a reasonable means of generating a comparable set of Hamiltonians with different geometries but similar time scales of interest.  See Fig. \ref{rotation_Er}.

Now, for the Hamiltonian in Eq. \ref{langevin_H}, we have yet to define the precise form of the constants $c_\alpha$ and $c_\beta$. For a Langevin model, we require each of the two baths to have spectral densities of the form $J(\omega)=\gamma\omega$, where $\gamma$ is the damping factor and defines the strength of the apparent friction on the primary modes. If $\rho_\alpha$ and $\rho_\beta$ are the densities of states in their respective baths, the values of the $c_\alpha$ and $c_\beta$ terms are defined via the spectral function as follows:
\begin{align}
    J_\alpha(\omega_\alpha)=\frac\pi2\frac{c_{\alpha}^2}{\omega_\alpha}\rho_\alpha(\omega_\alpha)=\gamma\omega_\alpha\quad\Rightarrow\quad c_{\alpha}=\sqrt{\frac2\pi\frac{\gamma\omega_\alpha^2}{\rho_\alpha(\omega_\alpha)}}\\
    J_\beta(\omega_\beta)=\frac\pi2\frac{c_{\beta}^2}{\omega_\beta}\rho_\beta(\omega_\beta)=\gamma\omega_\beta\quad\Rightarrow\quad c_{\beta}=\sqrt{\frac2\pi\frac{\gamma\omega_\beta^2}{\rho_\beta(\omega_\beta)}}  
\end{align}

If we let $\bm c_\alpha$ and $\bm c_\beta$ be the vectors of couplings $c_\alpha$ and $c_\beta$, and we let $\textbf q$ be the vector containing all the position operators of the system:
\begin{equation}
    \hat{\textbf q} = \begin{bmatrix}\hat x & \hat y & \{\hat q_\alpha\} &\{\hat q_\beta\} \end{bmatrix}^T
\end{equation}
then we can rewrite Hamiltonian in \ref{langevin_H} in the form of Eq. \ref{quad_H}:
\begin{equation}\label{langevin_H_transformed}
\mathcal{H} = \frac{\hat{\textbf p}\cdot\hat{\textbf p}}{2}+\begin{bmatrix}
        \frac12\hat{\textbf q}^T\hat{\bm\Omega}^2_g\hat{\textbf q}+\bm\lambda_g^T\hat{\textbf q}+\Tilde E_g & Ve^{i\textbf W^T\hat{\textbf q}}\\
        V^*e^{-i\textbf W^T\hat{\textbf q}}& \frac12\hat{\textbf q}^T\hat{\bm\Omega}^2_e\hat{\textbf q}+\bm\lambda_e^T\hat{\textbf q}+E_e
    \end{bmatrix}
\end{equation}
\begin{equation*}
    \hat{\bm{\Omega}}_g^2=\begin{bmatrix}
        \omega_2^2+\sum_\alpha \frac{c_\alpha^2}{\omega_\alpha^2} & 0 & \bm{c}_\alpha^T & \bm{0}\\
        0 & \omega^2_1+\sum_\beta\frac{c_\beta^2}{\omega_\beta^2} & \bm{0} & \bm{c}^T_\beta\\
        \bm{c}_\alpha & \bm{0} & \bm{\Omega}_\alpha^2 &\bm{0}\\
        \bm{0} & \bm{c}_\beta &\bm{0} & \bm{\Omega}_\beta^2
    \end{bmatrix}
\end{equation*}
\begin{equation*}
    \hat{\bm{\Omega}}_e^2=\begin{bmatrix}
        \omega_1^2\cos^2\phi+\omega_2^2\sin^2\phi + \sum_\alpha \frac{c_\alpha^2}{\omega_\alpha^2} & -\frac12\sin(2\phi)(\omega_2^2-\omega_1^2) & \bm{c}_\alpha^T & \bm{0}\\
        -\frac12\sin(2\phi)(\omega_2^2-\omega_1^2) & \omega^2_2\cos^2\phi+\omega_1^2\sin^2\phi+\sum_\beta\frac{c_\beta^2}{\omega_\beta^2} & \bm{0} & \bm{c}^T_\beta\\
        \bm{c}_\alpha & \bm{0} & {\bm{\Omega}}_\alpha^2 &\bm{0}\\
        \bm{0} & \bm{c}_\beta &\bm{0} & {\bm{\Omega}}_\beta^2
    \end{bmatrix}
\end{equation*}
\begin{equation*}
    \bm\lambda_e=\begin{bmatrix}
    0 & 0 & \bm 0 & \bm 0
    \end{bmatrix}^T\qquad \bm\lambda_g=-d\begin{bmatrix}
        \omega_2^2\cos\theta & \omega_1^2\sin\theta & \bm 0 & \bm 0
    \end{bmatrix}^T\qquad\textbf W=W\begin{bmatrix}
        \cos\eta &\sin\eta & \bm 0 & \bm 0
    \end{bmatrix}^T
\end{equation*}
\begin{equation*}
    \Tilde E_g=E_g+\frac12\left(\omega_1^2\sin^2\theta+\omega_2^2\cos^2\theta\right)\cdot d^2
\end{equation*}

From here, we simply apply the procedure outlined in Section \ref{dusc_mot} to obtain the ground- and excited-state normal mode matrices, the Duschinskii matrix $\textbf S$, and the shift vector $\textbf d$, and then evaluate nonequilibrium Fermi golden rule rates.

In principle, the Hamiltonian that has been defined is infinite dimensional; it is in this limit that we recover Langevin dynamics. In practice, however, it is obviously impossible to simulate an infinite-dimensional system. A brief note on the numerical implementation is provided in Appendix \ref{implementation}.

\section{Results}\label{Results}

The model above is characterized by quite a few parameters. A summary of what the various model parameters mean in the context of the system being studied is provided in Table \ref{constants} for the reader's convenience.

\begin{table}[h]
    \centering
    \begin{tabular}{|c|c|}
        \hline
         Parameter &  Physical meaning\\
         \hline
         $\gamma$ & The damping factor which defines the friction in the system\\
         \hline
         $V$ & The magnitude of the interstate coupling\\
         \hline
         $\phi$ & The rotation angle of the excited-state well orientation\\
         \hline
         $\theta$ & The rotation angle of the primary-mode shift vector\\
         \hline
         $d$ & The magnitude of the primary-mode shift vector\\
         \hline
         $\eta$ & The rotation angle of the spin-orbit vector \\
         \hline
         $W$ & The magnitude of the spin-orbit coupling vector\\
         \hline
         
    \end{tabular}
    \caption{A summary of the parameters that define the system at hand. The first three terms define the strengths of the various couplings in the system; the latter four define the physical geometry.}
    \label{constants}
\end{table}

All of the results below use the following parameters (atomic units):
\begin{multicols}{2}
    $$\omega_1 = 2\cdot10^{-4}$$
    $$\omega_2 = 4\cdot10^{-4}$$
    $$\gamma = 4\cdot10^{-4}$$
    
    $$W = 0.05$$
    $$\beta = 1000$$
    $$V = 1\cdot10^{-4}$$
\end{multicols}

There are three different values of $\theta$ investigated: $\pi/4$, $\pi/6$, and 0. For the first case, $d$ is set as 884 au; for the latter two cases, $d$ is set as 769 au.  (The value for $d$ has been adjusted for the latter two cases in order to keep the activation energy approximately constant as $\theta$ is changed.) Finally, three different values of $\Delta G$ are probed: 0, -0.01, and -0.029. Note that $\Delta G$ is not the bias between the primary mode wells, but rather the energy gap between the well minima after coupling to the bath is accounted for.  In other words, $\Delta G$ is as defined in Eq. \ref{H_B^g} (and is not the same as $E_g$ in Eq. \ref{langevin_H} or $\Tilde E_g$  in Eq. \ref{langevin_H_transformed}).  The rationale behind these choices for $\Delta G$ will be explained below.
    
\subsection{Equilibrium Dynamics: System Characterization}

\begin{figure}
    \centering
    \captionsetup{width=0.8\linewidth}
    \includegraphics[scale=0.5]{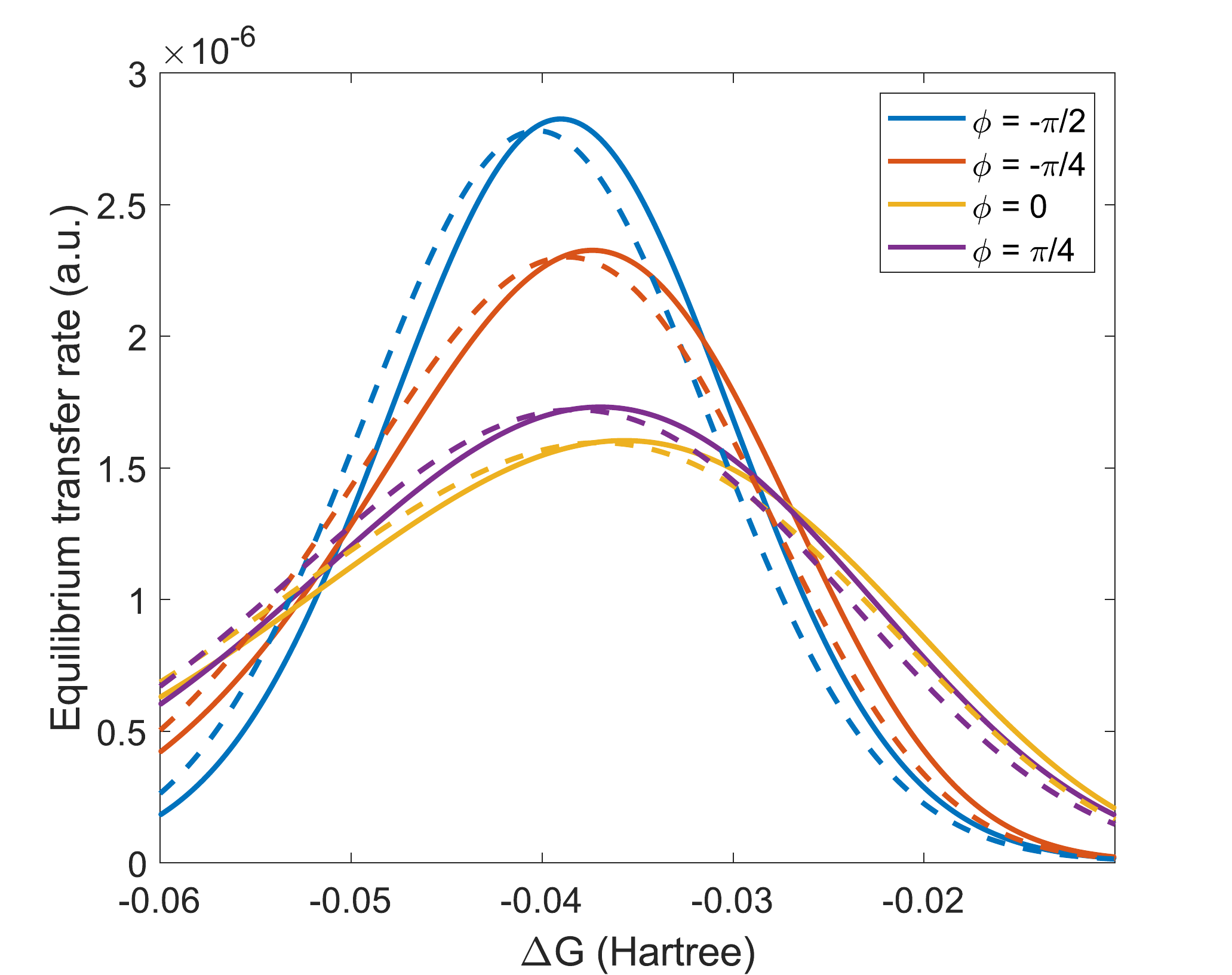}
    \caption{Equilibrium transfer rate plotted against $\Delta G$ for the model Hamiltonian in Eq. \ref{langevin_H} with $\theta=\pi/4$. Solid curves are for $W=0$ and dashed curves are for $W=0.05$ with $\eta=\pi/2$; in practice, the value of $\eta$ does not lead to to any significant effect here. Note also that changing the rotation angle $\phi$ does not significantly alter the $\Delta G$ at which the maximal transfer rate is achieved (i.e. the rate-defining reorganization energy of the system).}
    \label{marcus}
\end{figure}    

We begin with a brief look at the equilibrium behavior of the model as a means of characterizing the system. This characterization will be performed based on Marcus-style rate curves, where the equilibrium transfer rate is plotted as a function of $\Delta G$. Several of these curves for different values of $\phi$ are plotted in Fig. \ref{marcus}. Here, the well geometry used is $\theta=\pi/4$; the corresponding plots for the other 2 geometries are nearly identical. From these curves, we can extract information about the standard and geometric reorganization energies of the system.

First, note that the system has the desired property that changing the rotation angle $\phi$ does not significantly alter the electron transfer rates (which are on a similar order of magnitude).

Next, notice that for all rotation angles $\phi$, adding spin-orbit coupling to the system (making $W$ nonzero) slightly increases the activation barrier, shifting the Marcus curve to the left with a magnitude of about 0.0012 Hartree for all $\phi$. By analogy with standard Marcus theory, one might then argue that the ``effective'' reorgnization energy has therefore experienced a small shift, on the order of 3-4 per cent. 
This behavior is consistent with the behavior of the simpler shifted harmonic oscillator model seen in Ref. \cite{chandran:spinrates:2022} where one could easily define a total reorganization (without the headache of having two different reorganization energies from donor-to-acceptor and acceptor-to-donor).

\subsection{Photoexcitation Dynamics}

\begin{figure}[t!]
    \centering
    \includegraphics[scale=0.6]{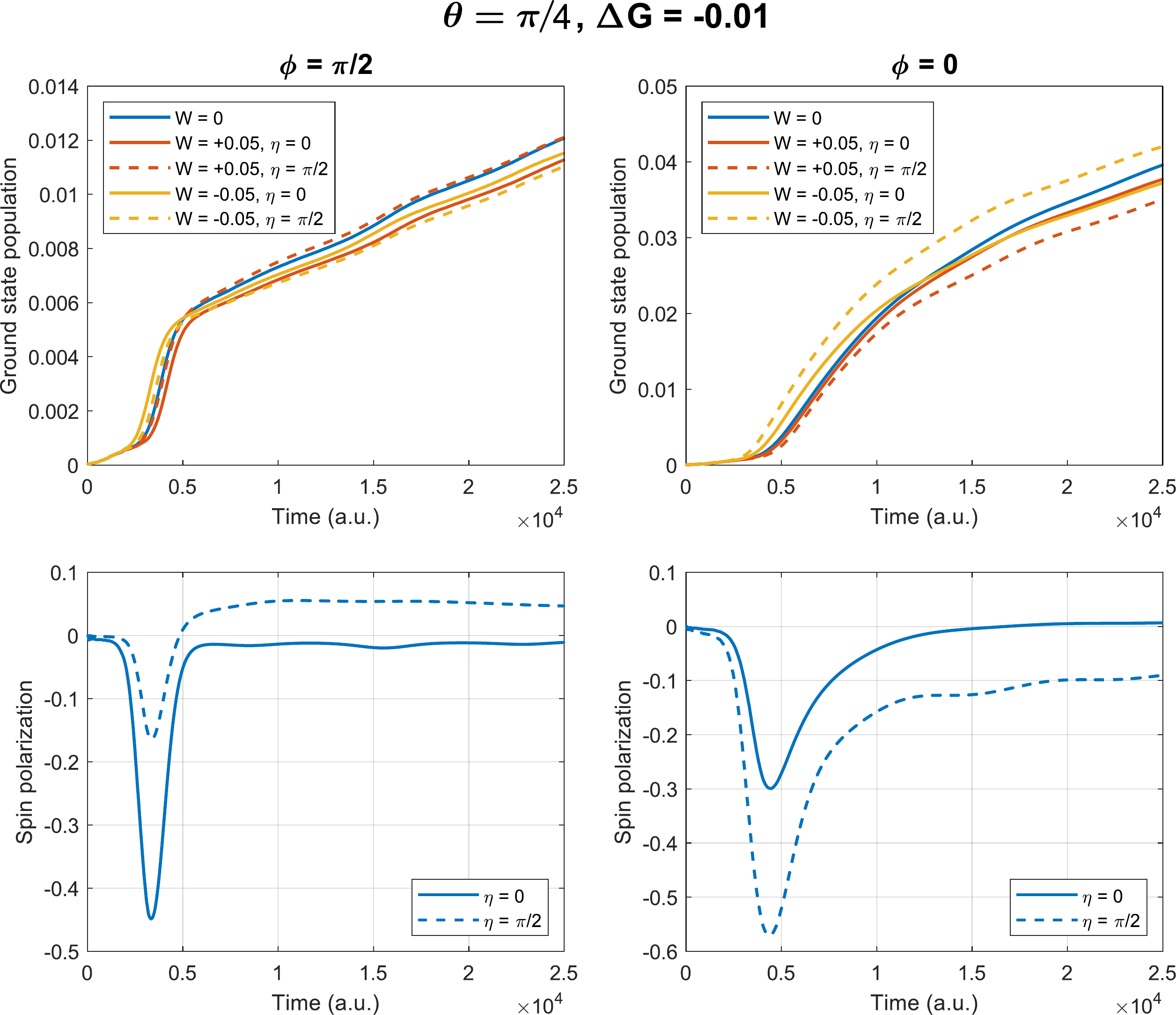}
    \caption{Plotted are ground state populations (top) and spin polarizations (bottom) following a fast photoexcitation for two pairs of systems. On the left are two SHO systems ($\phi = \pi/2$), and on the right are two systems which have a 90 degree well rotation ($\phi = 0$). In each panel, one of the systems has the spin-orbit coupling vector $\textbf W$ directed along the x-axis ($\eta=0$) and the other has has $\textbf W$ directed along the y-axis ($\eta = \frac\pi2$). Note that although the final  population averaged over the up- and down-spin dynamics is approximately the same for the two values of $\eta$ (i.e. the average of the dashed red and dashed yellow lines is near the average of the solid red and solid yellow lines), the spin polarization for $\eta = 0$ is very different from the spin polarization for $\eta = \pi/2$ (especially in the case $\phi = 0$).}
    \label{eta_comp}
\end{figure}

Before presenting the main results pertaining to spin polarization, let us establish a simple picture of the relaxation dynamics in our model. The ground state population curve can be loosely characterized by three epochs, which can be observed in the dynamics for any parameter set. First, there is a brief epoch of no population growth. Then, there is a large spike in the growth rate. Finally, the rate spike decays leaving a steady, constant growth of the ground state population. As an example, see Fig. \ref{eta_comp}.
These three epochs can be understood semiclassically as follows: First, following the photoexcitation, the nuclei are quite far from the crossing seam between the excited and ground state diabats. Therefore, as they begin to evolve on the excited state surface, not much population transfer can take place. However, the nuclei soon approach the crossing seam. Here, population is readily transferred from the excited state to the ground state. However, all the while, energy is being dissipated by the large bath coupled to the primary modes. This dissipation causes the system to decohere, ultimately resulting in a decay to equilibrium dynamics, which accounts for the steady and constant growth seen in the third epoch of the dynamics.
For our model, we find that 25,000 a.u. is sufficient time for the system to fully decohere; the absolute difference in ground state up- and down-spin populations will remain constant after this time. Because we are primarily interested in the persistence of spin-polarization, we will perform all polarization analysis at this time, after decoherence has completed.

\begin{figure}[t!]
    \centering
    \includegraphics[scale=0.65]{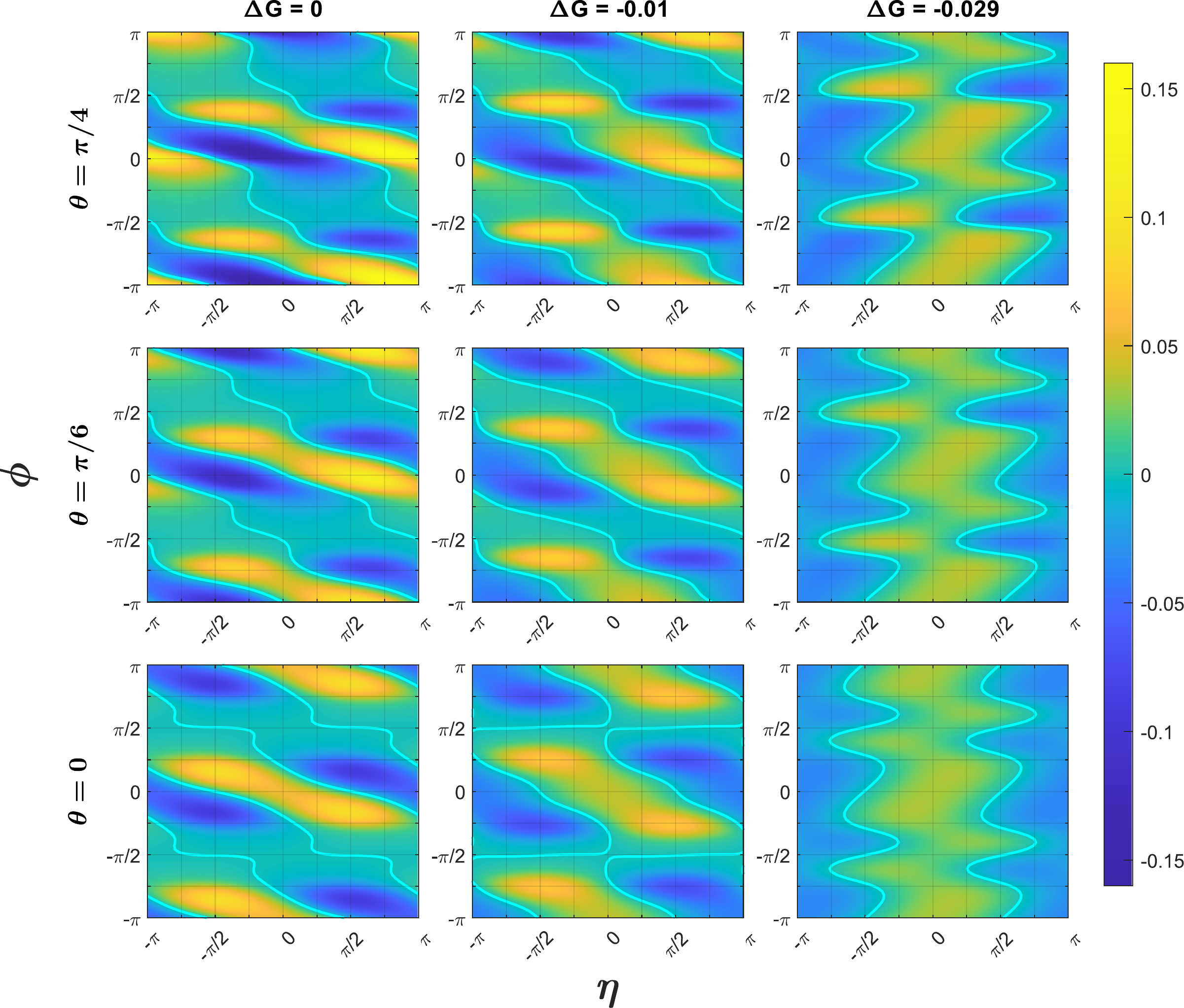}
    \caption{Plotted are the final ground-state spin polarizations ($\chi$ in Eq. \ref{pol_def}) as functions of $\phi$ and $\eta$ for three different values of $\theta$ and three different values of $\Delta G$. The three columns from left to right correspond respectively to $\Delta G=0$, $\Delta G=-0.01$, and $\Delta G=-0.029$. The three rows from top to bottom respectively correspond to $\theta=\pi/4$, $\theta = \pi/6$, and $\theta = 0$ (with respective values of $d=884$, $d=769$, and $d=769$).  Also plotted in solid cyan are the zero-polarization isolines. Note that the polarization can depend quite sensitively on the values of $\phi$ and $\eta$ (i.e. the geometry of the system).}
    \label{pols_pops}
\end{figure}

\begin{figure}[t!]
    \centering
    \includegraphics[scale=0.65]{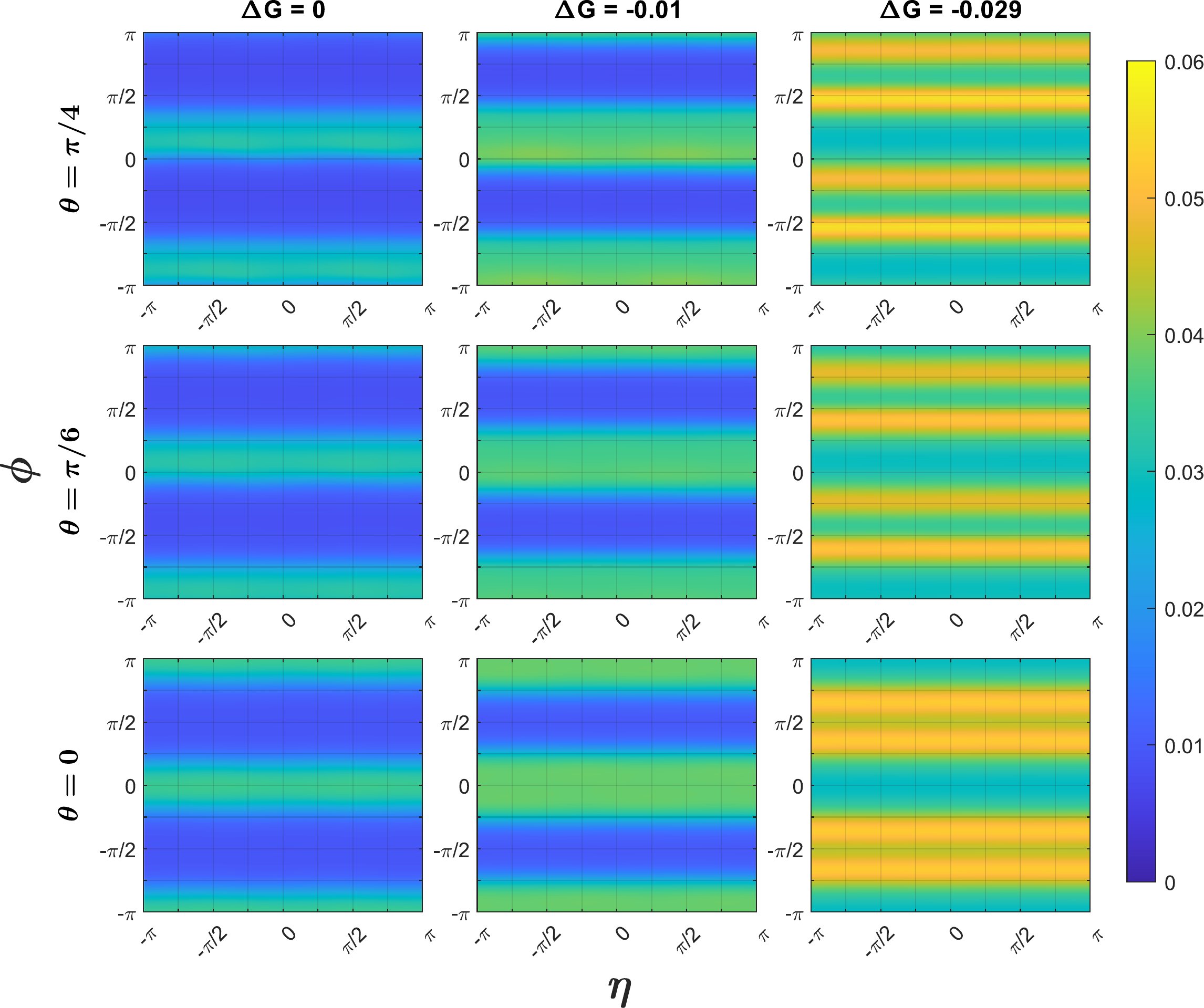}
    \caption{Plotted are the final ground-state populations $P_g$ as functions of $\phi$ and $\eta$ for three different values of $\theta$  and three different values of $\Delta G$. Each panel uses the same parameter set as the corresponding panel in Fig. \ref{pols_pops}. Note that the population can vary greatly with $\phi$, but shows very little dependence on $\eta$.}
    \label{pols_pops2}
\end{figure}

In  Figs. \ref{pols_pops} and \ref{pols_pops2}, we plot the 
polarization and the ground state populations at time 25,000 au. If we denote the ground-state populations for spin-up and spin-down as $P^+$ and $P^-$, the polarization is defined as:
\begin{equation}\label{pol_def}
  \chi =   \frac{P^+-P^-}{P^++P^-}
\end{equation}
Similarly, the total ground state population is defined follows:
\begin{equation}\label{pop}
    P_g = \frac{P^++P^-}{2}
\end{equation}
Note that we expect a realistic initial state to be completely unpolarized, i.e. the initial electronic populations should be equally split between spin-up and spin-down. Therefore within our model, because we  ignore spin-flip processes, one can calculate the total overall population dynamics (with Eq. \ref{pop}) by taking a simple average of the spin-up and spin-down populations.

For both of these data sets (Fig. \ref{pols_pops} and \ref{pols_pops2}), dynamics were computed with different parameter pairs ($\phi$, $\eta$), different well positions ($d$, $\theta$), and a $\Delta G$ of either 0, -0.01 or -0.029.  These three regimes for $\Delta G$ were chosen based on the Marcus curves in Fig. \ref{marcus}. On the one hand, -0.01 Hartree is a modest $\Delta G$ for which there is a nonzero donor-acceptor driving force but still a large activation barrier; the equilibrium transfer rate is quite sensitive to $\phi$.  On the other hand, -0.029 is a value for $\Delta G$ at which all the rate curves for different values of $\phi$ nearly intersect, suggesting similarity in activation barrier heights. Finally, at $\Delta G = 0$, the system is exactly symmetric in ground- and excited-state well heights. 

Note several general features of the data in Fig. \ref{pols_pops}. First, note that the data is $\pi$-periodic in $\phi$ and $\pi$-antiperiodic in $\eta$. The periodicity in $\phi$ is easily explained by the fact that the well geometry for $\phi=\phi_0$ is exactly the same as the well geometry for $\phi=\phi_0+n\pi$. The $\eta$ antiperiodicity can be understood from the simple observation that:
\begin{equation}
    W(\hat x\cos\eta+\hat y\sin\eta)=-W(\hat x\cos(\eta+\pi)+\hat y\sin(\eta+\pi))
\end{equation}

Second, by comparison with  Fig. \ref{pols_pops2}, note that $\eta$ (which, as a reminder, defines the direction of the spin-orbit coupling vector) can dramatically change the spin polarization of the final state without significantly altering the final population $P_g$. This finding indicates that the observed spin polarization effects do not necessarily correlate with the final state populations.  This clear distinction  between polarization and population  is highlighted in Fig. \ref{eta_comp}, where we plot dynamics for two different pairs of systems, where the only difference within each pair is the direction of the $\textbf W$ vector.  

\begin{figure}
    \centering
    \includegraphics[scale=0.6]{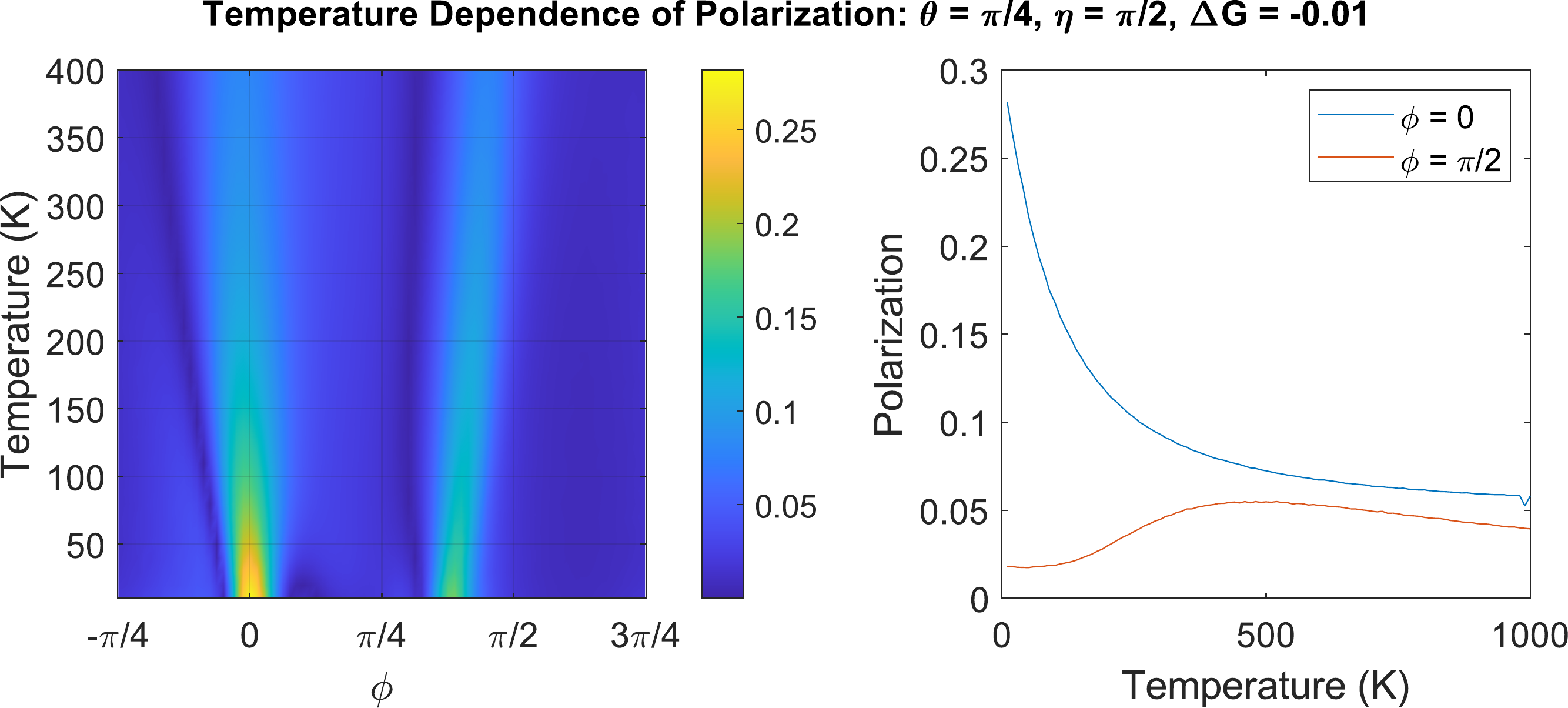}
    \caption{Plotted on the left is the magnitude of the final ground state spin-polarization as a function of $\phi$ and temperature for the configuration $\theta=\pi/4$, $\eta=\pi/2$, $\Delta G=-0.01$. Note that there are very specific values of $\phi$ which promote spin polarization at low temperatures. In particular, when $\phi=0$, spin-polarization  increases monotonically as temperature tends to 0. This trend is in stark contrast to the $\phi=\pi/2$ (shifted harmonic oscillator) case, in which the spin-polarization attains a maximum  around 400 K.}
    \label{temp_dependence}
\end{figure}

Finally, in Fig. \ref{temp_dependence}, we investigate the temperature-dependence of the spin-polarization effects observed above. We find that the well rotation angle $\phi$ is tightly coupled to the low-temperature behavior of the spin polarization. As an example, notice in Fig. \ref{temp_dependence} that for the shifted harmonic oscillator case $\phi=\pi/2$, the polarization attains a maximum at around 400 K, whereas adding a Duschinskii rotation and switching to the $\phi=0$ case results in a completely monotonic dependence on temperature.

\section{Discussion}\label{Discussion}

\begin{figure}[t!]
    \centering
    \captionsetup{width=.83\linewidth}
    \includegraphics[scale=0.4]{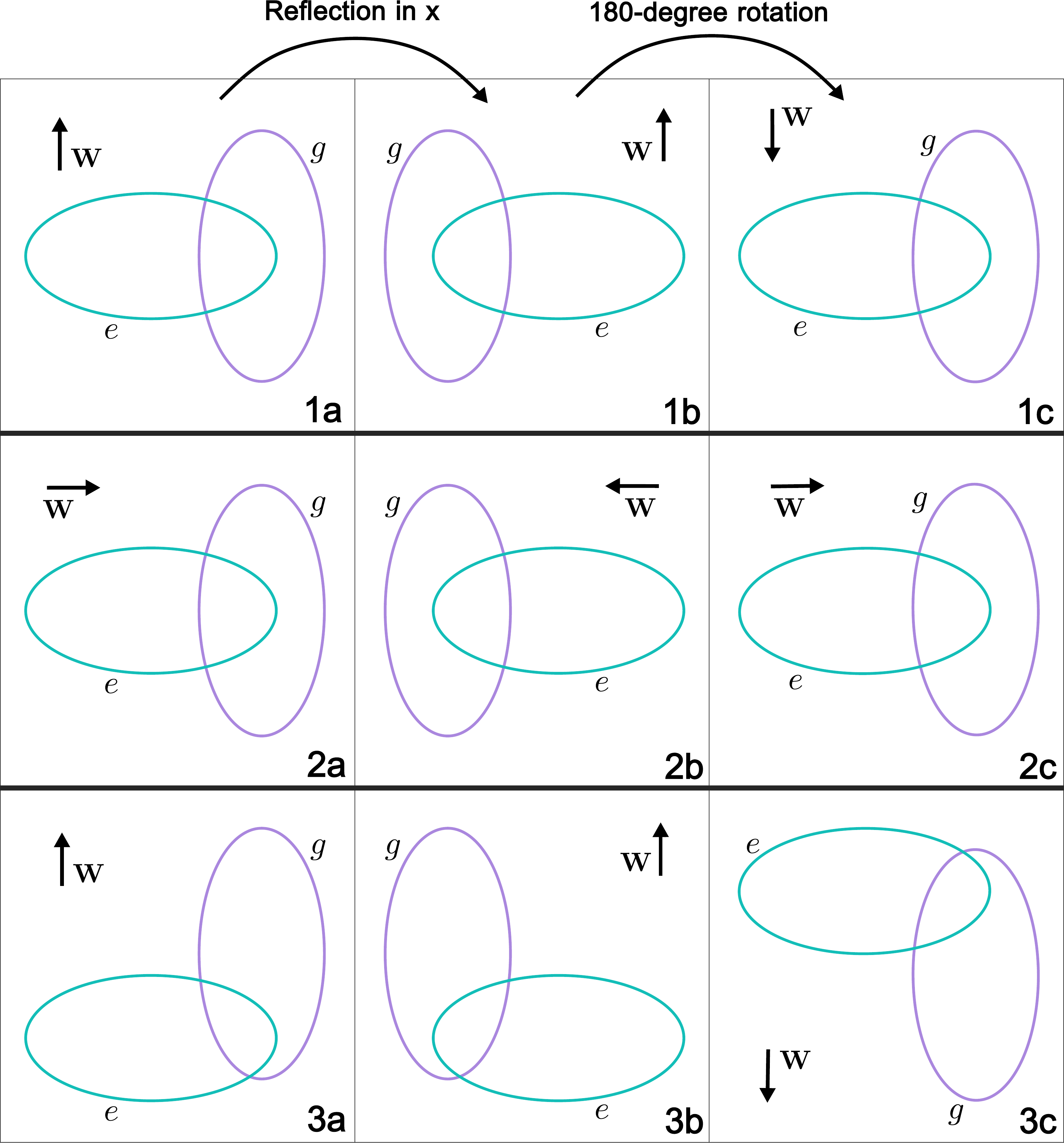}
    \caption{The dynamics in the \textbf{a} panels are the same as those in the corresponding \textbf{b} and \textbf{c} panels. This is due to the Hamiltonian in Eq. \ref{langevin_H_transformed} being invariant under reflections and rotations in position space. This allows us to easily connect chirality to the emergence of spin-dependent transfer effects. On the one hand, the dynamics of system $\bm 1\textbf{a}$ will have no spin-dependence; system $\bm 1\textbf{c}$ is precisely the $-\textbf W$ analog of system $\bm 1\textbf{a}$. On the other hand, the dynamics of systems $\bm 2\textbf{a}$ and $\bm 3\textbf{a}$ are not expected to exhibit spin-independence, since systems $\bm 2\textbf{c}$ and $\bm 3\textbf{c}$ are not superimposable on the $-\textbf W$ analogs of $\bm 2\textbf a$ and $\bm 3\textbf a$. Although system $\bm 2\textbf a$ is spatially achiral, it possesses asymmetry with respect to $\textbf W$. System $\bm 3\textbf{a}$ has spatial chirality, so any spatial reflection will lead to a nonsuperimposable system.}
    \label{achiral_wells}
\end{figure}

Overall, the observations above allow us to readily answer both of the two overarching questions posed in the introduction. First, the direction of the spin-orbit coupling vector is extremely important in generating spin polarization in electron transfer, and the optimal direction is tightly coupled to the geometry of the system. Second, changes in normal mode frequencies and couplings upon electron transfer can greatly promote spin polarization over shifted harmonic oscillator systems, and these effects can be amplified significantly at low temperatures. As can be seen in Fig. \ref{pols_pops}, significantly more spin polarization can be observed in systems with a 90 degree well rotation ($\phi=0$) over the SHO case ($\phi=\frac\pi2$). Additionally, we see that $\phi=0$ systems more consistently produce sizable spin polarization than $\phi=\frac\pi2$ systems. Although we have been able to produce a satisfying answer to the two overarching questions of this investigation, we would also ultimately like some intuition as to the general shape of the polarization landscape. Despite the evident complexity, as well as the nontrivial relationship the landscape has with $\Delta G$, we will put forward a hypothesis that at least partially explains the observed results.

Consider the various systems depicted in Fig. \ref{achiral_wells}. The electronic dynamics of the systems in the $\textbf a$ panels are exactly the same as the dynamics in the corresponding $\textbf b$ and $\textbf c$ systems. In order to see this equivalence, note that the $\textbf b$ systems are obtained by mirror reflecting the left-hand systems in space, and that the $\textbf c$ panels are subsequently obtained by a simple rotation of the $\textbf b$ systems. The electronic dynamics of the Hamiltonian in \ref{langevin_H_transformed} are invariant to such transformations, which are equivalent to basis transformations of the nuclear coordinates. This relationship demonstrates the natural connection between the spin-dependent effects observed above and system chirality. We can easily see that system $\bm 1\textbf a$ in Fig. \ref{achiral_wells} must have exactly the same dynamics for $\pm\textbf W$, since system $\bm 1\textbf c$ is exactly superimposable onto its $-\textbf W$ analog. A chiral system (such as that in panel $\bm 3\textbf a$) lacks spatial symmetry, and will typically exhibit dynamical differences for different spin states. In such a system, it is not possible to map $+\textbf W$ dynamics to $-\textbf W$ dynamics through operations which respect the symmetry of the Hamiltonian (i.e. spatial reflections and rotations).

However, in the systems presently under consideration, spatial symmetry is not enough to eliminate spin-dependent transfer effects; consider system $\bm 2\textbf a$ in Fig. \ref{achiral_wells}. Although the potential wells are superimposable upon one another following a rotation, we find that the $\textbf W$ vector of $\bm 2\textbf c$ does not map to $-\textbf W$ in $\bm 2\textbf a$. Consequently, there is no reason to expect that the dynamics of system $\bm 2\textbf a$ will be spin-independent. Therefore, for the purposes of understanding spin-dependent dynamics, we must adopt a much more restrictive notion of achirality than is standard; for our purposes, an achiral system must satisy two constraints. First, as usual, $(i)$ there must be a mirror plane; second, not as usual, $(ii)$ that mirror plane must be  {\em orthogonal to} $\textbf W$. Only in such a case, when the reflection maps $\textbf W$ to $-\textbf W$, will we be left with a system that is spatially superimposable onto the original system.

\begin{figure}[t!]
    \centering
    \includegraphics[scale=0.46]{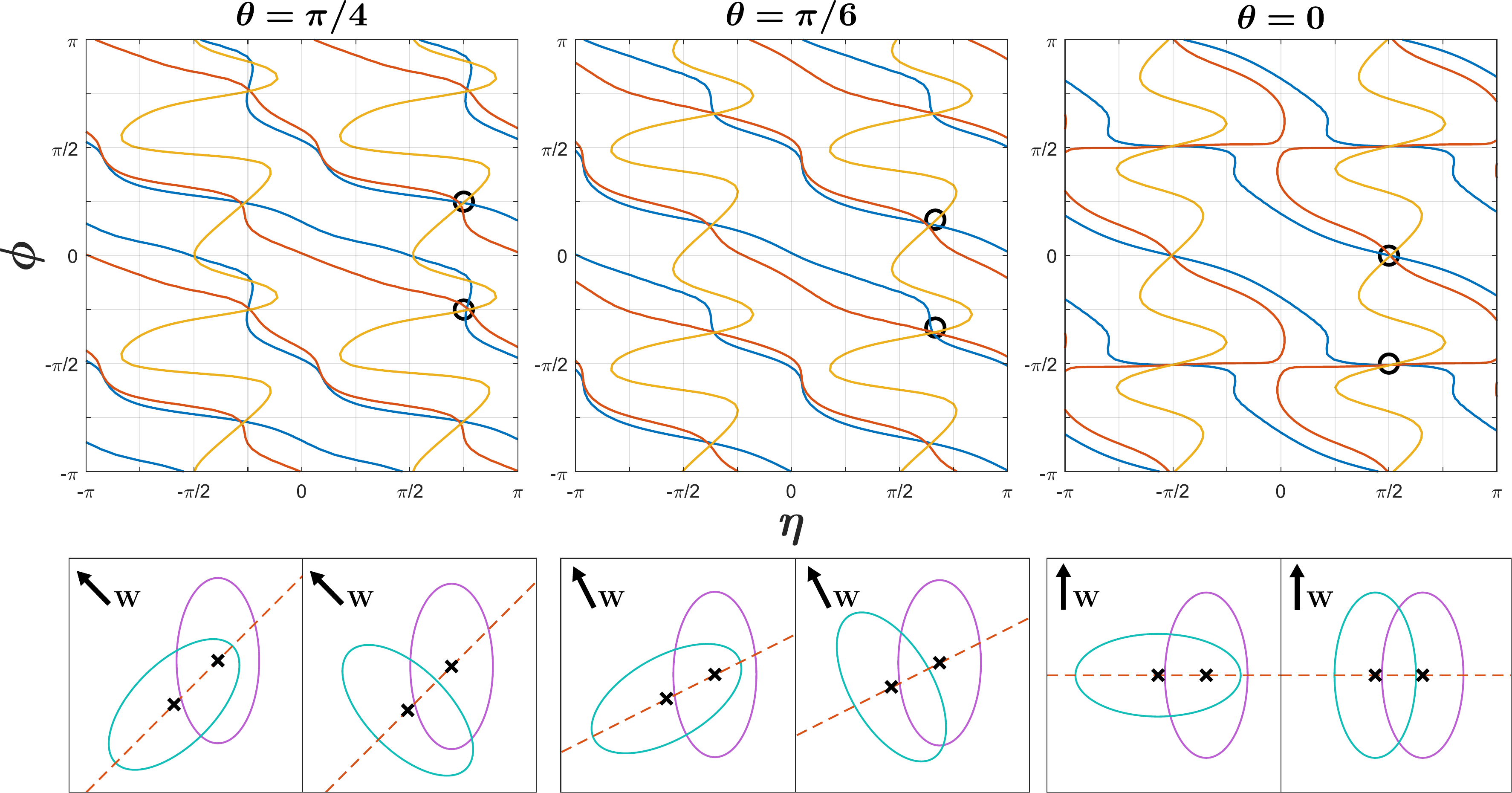}
    \caption{(Upper) The three zero-polarization isolines for each well positioning are plotted together. Blue, yellow, and red traces correspond respectively to $\Delta G$ values of 0, -0.01, and -0.029. Black circles mark the points $(\phi,\eta)=\left(\pm\frac\pi4,\frac{3\pi}4\right)$ on the left, $(\phi,\eta)=\left(\frac\pi6,\frac{2\pi}3\right),\left(-\frac\pi3,\frac{2\pi}3\right)$ in the middle, and $(\phi,\eta)=\left(0,\frac{\pi}2\right),\left(-\frac\pi2,\frac{\pi}2\right)$ on the right. Note that, in every panel, all three isolines intersect very close to these target points which designate expected minimal chirality configurations. (Lower) Pictured for each value of $\theta$ are the two minimum chirality configurations designated by the black circles. In each panel, the ground state (purple) minimum is indicated with an ``x'' and the relevant reflection plane (i.e. that plane orthogonal to $\textbf W$ which bisects the excited state well) is indicated with a red dashed line.}
    \label{isolines}
\end{figure}

Now we return to the systems investigated in Fig. \ref{pols_pops}. We expect minimum polarizations to occur at some system configurations where chirality (as per the above definition) is ``minimized''; in other words, the system is as close to achiral as possible. First, we focus on the case $\theta=\pi/4$. In the spirit of condition $(i)$, we will first identify values of $\phi$ which produce well configurations with relatively high spatial symmetry. $\phi=\frac{\pi}4$ and $\phi=-\frac\pi4$ are two such configurations; in each of these two systems, the major or minor axis of the excited state surface is oriented along the axis connecting the well minima (see Fig. \ref{isolines}, where the relevant configurations are pictured in the lower panels). Next, following condition $(ii)$, we expect that the minimum chirality configuration will be achieved with $\eta=\frac{3\pi}{4}$, because this condition places the mirror plane on the axis connecting the well minima, making the excited state surface exactly symmetric (again, see Fig. \ref{isolines}). 
These intuitive arguments are indeed confirmed by the numerical results in Fig. \ref{pols_pops}. In Fig. \ref{isolines}, we take the zero-polarization isolines from Fig. \ref{pols_pops} and plot them together for each value of $\theta$ (i.e. each variation of well position). We see that, even as $\Delta G$ changes, the $\theta=\pi/4$ zero-polarization level curves pass nearly exactly through both of the points $(\phi,\eta)=\left(\frac\pi4,\frac{3\pi}4\right), \left(-\frac\pi4,\frac{3\pi}4\right)$ (marked in Fig. \ref{isolines} with black circles). Similarly, the level curves for the case $\theta=\pi/6$ all nearly pass through the points $(\phi,\eta)=\left(\frac\pi6,\frac{2\pi}3\right),\left(-\frac\pi3,\frac{2\pi}3\right)$ and those for the case $\theta=0$ nearly pass through $(\phi,\eta)=\left(0,\frac\pi2\right),\left(-\frac\pi2,\frac{\pi}2\right)$. 

Interestingly, the notion of achirality developed above can also be applied (roughly) to understand the locations of the ``islands'' of strong polarization in Fig. \ref{pols_pops}. Consider again, for illustrative purposes, the case $\theta=\pi/4$. If minimum chirality configurations occur at $\phi=\pm\pi/4$, then we would naturally expect $\phi=0,\pi/2$ to produce {\em maximum} chirality configurations and have the potential for producing large spin polarization. This is again confirmed numerically in Fig. \ref{pols_pops}, where we see strong polarization islands around $\phi=0$ and $\phi=\frac\pi2$. Similarly, for $\theta=\pi/6$ and $\theta=0$ respectively, we observe islands of strong polarization around $\phi=\frac{5\pi}{12},-\frac{\pi}{12}$ and $\phi=\pm\frac\pi4$.  All of these observations taken together suggest that the strong
notion of chirality developed above may partly account for the polarization landscape in Fig. \ref{pols_pops}. However, if chirality were the entire story, we would expect the zero-polarization level curves to be independent of $\Delta G$, which is clearly not the case.
One must suppose that, for different values of $\Delta G$, the dynamics become more complicated as one cannot ignore the time scale for energy dissipation when calculating transient spin polarization.

As a final remark, it is interesting to note that Duschinskii rotations, in addition to promoting spin polarization, also appear to produce greater temporal separation between up- and down-spin dynamics (i.e. greater transient spin polarization). In other words, as can be seen in Fig. \ref{eta_comp}, the differences in populations for the up- and down-spin dynamics in the $\phi=0$ case are larger than they are for the $\phi=\pi/2$ (SHO) case; in fact, for both the $\eta=0$ and $\eta=\pi/2$ cases, the spin up and spin down populations largely overlap for the $\phi=\pi/2$ case. Notably, as $\eta$ varies,  Fig. \ref{eta_comp} shows that greater transient polarization at early times in the SHO model correlates with decreased final spin-polarization, whereas greater transient polarization at early times in the rotated-well model correlates with increased final spin polarization. Overall, this result suggests that if one can drive  nonequilibrium electron transfer dynamics with Duschinskii rotations so as to prevent equilibration, one might recover even larger spin polarizations than what we have found here. Future work may well investigate the effect of such a driving force.

\section{Conclusions}\label{Conclusions}

In conclusion, we have used a Duschinskii rotation formalism to investigate spin-polarization in a Langevin-ESOC model. Our investigation demonstrates that the final spin polarization can be quite large and highly sensitive to the exact system geometry. This finding has particular relevance for systems where changes in electronic state may be accompanied by significant changes in coupling between vibrational modes. A notion of chirality has been shown to partially explain how polarization can depend on system geometry, but further work is needed to understand and predict such polarization in general. While the model here is very abstract (and we have not yet used ab initio calculations to extract parameters for a given molecule), our conclusions highlight that standard Marcus models for electron transfer using only the donor and acceptor minimum geometries may be very inadequate as far as reproducing spin polarization of a realistic system. Future extensions of this work would also benefit by treating multi-electron problems.  For instance,  although a great deal of work has been done treating the singlet to triplet ISC problem, \cite{varganov:2016:ijqc,varganov:review:2022}, the spin polarization of product states has not yet been fully explored\cite{Fay2021,Limmer2021}.
In the future, it will also be very helpful to extend the two-level results presented here to the case of a metal surface with many electronic states, where the CISS phenomenon is often observed.\cite{naaman:2012:jpcl}

\section{Acknowledgements}
This work was supported by the U.S. Air Force Office of Scientific Research (USAFOSR), AFOSR Grant Nos. FA9550-18-1-0 497 and FA9550-18-1-0 420.

\section{Appendix}
In this appendix we provide details on the derivation of the equilibrium and nonequilibrium time correlation functions for a harmonic system incorporating Duschinskii rotations. We will make extensive use of two identities:
\begin{align}
    \int_{\mathbb{R}^n}dx^n\,\exp\left(-\frac12\textbf x^{\text{T}}\hat A\textbf x+B^{\text{T}}\textbf x\right)=\sqrt{\frac{(2\pi)^n}{\text{det}A}}\exp\left(\frac12B^T\hat A^{-1}B\right)\label{multi_gaussian}\\
    \Braket{\textbf x}{e^{-i\hat H t}}{\textbf y}=\sqrt{\frac{\det{\bm a(t)}}{(2\pi i)^N}}\exp\left(\frac i2\left[\bm x^T\bm b(t)\bm x+\bm y^T\bm b(t)\bm y-2\bm x^T\bm a(t)\bm y\right]\right)\label{HO_pathint}
\end{align}

The former identity is simply a multivariate Gaussian integral. The latter is derived from the explicit evaluation of the Green's function for a harmonic Hamiltonian $\hat H$ composed of $N$ independent harmonic modes. Here, $\bm a(t)=\hat{\bm \Omega}[\sin(\hat{\bm \Omega}t)]^{-1}$ and $\bm b(t)=\hat{\bm \Omega}[\tan(\hat{\bm \Omega}t)]^{-1}$, where $\hat{\bm \Omega}$ is the (diagonal) harmonic matrix of $\hat H$.

The Hamiltonian of interest is given in Eq. \ref{duschinskii_H}. We will rewrite it here for convenience:
\begin{equation}
    \hat H = \left(E_g+\hat H_B^{(g)}\right)\ket g\bra g+\left(E_e+\hat H_B^{(e)}\right)\ket e\bra e+Ve^{i\textbf W^T\hat{\textbf x}}\ket g\bra e+V^*e^{-i\textbf W^T\hat{\textbf x}}\ket e\bra g
\end{equation}
$$\hat H_B^{(e)} = \frac{\hat{\textbf p}\cdot\hat{\textbf p}}2+\frac12\hat{\textbf x}^\dagger\hat{\bm{\Omega}_e}^2\hat{\textbf x}$$
$$\hat H_B^{(g)} = \frac{\hat{\textbf p}\cdot\hat{\textbf p}}2+\frac12\left(\textbf S\hat{\textbf x}+\textbf d\right)^\dagger\hat{\bm{\Omega}_g}^2\left(\textbf{S}\hat{\textbf x}+\textbf d\right)$$
\subsection{Derivation of Eq. \ref{eq_corr}}\label{app_eq}

We start the system in the excited state with the density matrix:
\begin{align}
    \hat \rho(0) &= \frac{1}Z\exp(-\beta \hat H_B^{(e)})\ket e\bra e\\ 
    Z &= \text{Tr}\left[\exp(-\beta\hat H_B^{(e)})\right]
\end{align}

We want to approximate:
\begin{align}
    P_g(t)&=\frac{1}{Z}\text{Tr}_B\left[\Braket*{g}{e^{-i\hat Ht}e^{-\beta\hat H_B^{(e)}}}{e}\Braket*{e}{e^{i\hat Ht}}{g}\right]\\
    &=\frac{1}{Z}\text{Tr}_B\left[\Braket*{g}{e^{-i\hat Ht}}{e}e^{-\beta\hat H_B^{(e)}}\Braket*{e}{e^{i\hat Ht}}{g}\right]    
\end{align}

We make the following definition for convenience:
\begin{align}
    \hat H_0 &= \left(E_g+\hat H_B^{(g)}\right)\ket g\bra g+\left(E_e+\hat H_B^{(e)}\right)\ket e\bra e\\
    \hat V &= Ve^{i\textbf W^T\hat{\textbf x}}\ket g\bra e+V^*e^{-i\textbf W^T\hat{\textbf x}}\ket e\bra g
\end{align}
and expand $e^{-i\hat H t}$ in a Dyson series:
\begin{align}
e^{-i\hat Ht}&\simeq e^{-i\hat H_0t}\left(\textbf{I}-i\int_0^tdt''\,e^{i\hat H_0t''}\hat Ve^{-i\hat H_0t''}\right)\\
\Rightarrow\Braket*{g}{e^{-i\hat Ht}}{e}&\simeq-iVe^{-iE_gt}\int_0^tdt''\,e^{i(E_g-E_e)t''}e^{i\hat H_B^{(g)}t''}e^{i\textbf W^T\hat{\textbf x}}e^{-i\hat H_B^{(e)}t''}\\
\Rightarrow\Braket*{e}{e^{i\hat Ht}}{g}&\simeq iVe^{iE_gt}\int_0^tdt'\,e^{-i(E_g-E_e)t'}e^{-i\hat H_B^{(g)}t'}e^{-i\textbf W^T\hat{\textbf x}}e^{i\hat H_B^{(e)}t'}
\end{align}

\begin{align}
    \Rightarrow P_g(t)&\simeq\frac{\lvert V\rvert^2}{Z}\int_0^tdt'\,\int_0^tdt''\,e^{-i(E_g-E_e)(t'-t'')}\text{Tr}_B\left[e^{-(\beta-i(t'-t''))\hat H_B^{(e)}}e^{-i\textbf W^T\hat{\textbf x}}e^{-i\hat H_B^{(g)}(t'-t'')}e^{i\textbf W^T \hat{\textbf x}}\right]\\
    & =\frac{2\lvert V\rvert^2}{Z}\text{Re}\left[\int_0^tdt'\,\int_0^{t'}d\tau\,e^{-i(E_g-E_e)\tau}\text{Tr}_B\left[e^{-(\beta-i\tau)\hat H_B^{(e)}}e^{-i\textbf W^T\hat{\textbf x}}e^{-i\hat H_B^{(g)}\tau}e^{i\textbf W^T \hat{\textbf x}}\right]\right]
\end{align}

Letting $\tau_e = -\tau-i\beta$ and $\tau_g=\tau$, we need to compute the correlation function
\begin{equation}
    C(\tau)=\text{Tr}_B\left[e^{-i\hat H_B^{(e)}\tau_e}e^{-i\textbf W^T\hat{\textbf x}}e^{-i\hat H_B^{(g)}\tau_g}e^{i\textbf W^T \hat{\textbf x}}\right]
\end{equation}

We proceed by evaluating the trace using position coordinates:
\begin{align}
    C(\tau)&=\int_{\mathbb{R}^n}dx^n\,\Braket*{\textbf x}{e^{-i\tau_e\hat H_B^{(e)}}e^{-i\textbf W^T\hat{\textbf x}}e^{-i\hat H_B^{(g)}\tau_g}e^{i\textbf W^T \hat{\textbf x}}}{\textbf x}\\
    &=\int_{\mathbb{R}^n}dx^n\,\int_{\mathbb{R}^n}dy^n\,\int_{\mathbb{R}^n}dx'^n\,\int_{\mathbb{R}^n}dy'^n\,\Braket*{\textbf x}{e^{-i\tau_e\hat H_B^{(e)}}e^{-i\textbf W^T\hat{\textbf x}}}{\textbf y}\braket*{\textbf y}{\textbf y'}\Braket*{\textbf y'}{e^{-i\hat H_B^{(g)}\tau_g}}{\textbf x'}\Braket*{\textbf x'}{e^{i\textbf W^T \hat{\textbf x}}}{\textbf x}\label{position_basis_expansion}
\end{align}
$\textbf x'$ and $\textbf y'$ represent ground state coordinates, while $\textbf x$ and $\textbf y$ represent excited state position coordinates. Therefore, we have that $\braket{\textbf y}{\textbf y'}=\delta(\bm y'-(\textbf S\bm y+\textbf d))$. Using this fact and Eq. \ref{HO_pathint}, the integral reduces to
\begin{align}
    \sqrt{\frac{\det(\bm a_e)}{(2\pi i)^n}}\sqrt{\frac{\det(\bm a_g)}{(2\pi i)^n}}\int_{\mathbb R^n}dx^n\,\int_{\mathbb R^n}dy^n\,e^{i\left[\frac12\left(\bm x^{\text{T}}\bm b_e\bm x+\bm y^{\text{T}}\bm b_e\bm y\right)-\bm x^{\text{T}}\bm a_e\bm y\right]}\times\label{corrfn_gaussian}\\
    e^{i\left[\frac12\left(\left(\textbf S\bm x+\textbf d\right)^{\text{T}}\bm b_g\left(\textbf S\bm x+\textbf d\right)+\left(\textbf S\bm y+\textbf d\right)^{\text{T}}\bm b_g\left(\textbf S\bm y+\textbf d\right)\right)-\left(\textbf S\bm x+\textbf d\right)^{\text{T}}\bm a_g\left(\textbf S\bm y+\textbf d\right)\right]}e^{i\textbf W^T\bm x}e^{-i\textbf W^T\bm y}\nonumber
\end{align}

Here, $\bm a_e$ and $\bm a_g$ are shorthand for $\bm a_e(\tau_e)$ and $\bm a_g(\tau_g)$, and these functions are as in Sec. \ref{Dusc_dynamics}. Next, we focus on the integrand. We would like to collect terms and put it in the form of a multivariate Gaussian. Recall the following operators from Sec. \ref{Dusc_dynamics} (again using the shorthand in Eq. \ref{corrfn_gaussian}):
\begin{align}
    \textbf{A} &= \bm a_e+\textbf S^T\bm a_g\textbf S\\
    \textbf{B} &= \bm b_e+ \textbf S^T\bm b_g\textbf S\\
    \textbf{G} &= \bm b_g-\bm a_g\\
    \textbf{E} &= \bm b_e-\bm a_e\\
\end{align}

We can use these operators to write the integrand as
\begin{align}
    \exp\left(i\left[\frac12\left(\bm x^T\textbf B\bm x+\bm y^T\textbf B\bm y\right)-\bm x^T\textbf A\bm y+\left(\textbf d^T\textbf{GS}+\textbf W^T\right)\bm x+\left(\textbf d^T\textbf{GS}-\textbf W^T\right)\bm y+\textbf d^T\textbf G\textbf d\right]\right)\\
    =\exp\left(i\left[\frac12\begin{bmatrix}
    \bm x^T & \bm y^T
    \end{bmatrix}\begin{bmatrix}
    \textbf B & -\textbf A\\
    -\textbf A & \textbf B
    \end{bmatrix}\begin{bmatrix}
    \bm x\\
    \bm y\\
    \end{bmatrix}+\begin{bmatrix}
    \textbf d^T\textbf{GS}+\textbf W^T & \textbf d^T\textbf{GS}-\textbf W^T
    \end{bmatrix}\begin{bmatrix}
    \bm x\\
    \bm y\\
    \end{bmatrix}+\textbf d^T\textbf G\textbf d\right]\right)
\end{align}

Thus, the double integral above becomes a single integral over $\mathbb{R}^{2n}$:
\begin{equation}
    \int_{\mathbb{R}^{2n}}dz^{2n}\,\exp\left(-\frac12\left(\bm z^T\begin{bmatrix}
    -i\textbf B & i\textbf A\\
    i\textbf A & -i\textbf B
    \end{bmatrix}\bm z\right)+i\begin{bmatrix}
    \textbf d^T\textbf{GS}+\textbf W^T & \textbf d^T\textbf{GS}-\textbf W^T
    \end{bmatrix}\bm z+i\textbf d^T\textbf G\textbf d\right)
\end{equation}
where ${\bm z} = \left[\begin{array}{c} {\bm x} \\ {\bm y} \end{array} \right].$

Using the multivariate Gaussian integral identity, this evaluates to
\begin{equation}\label{intermediate_form}
    \sqrt{\frac{(2\pi)^{2n}}{(-i)^{2n}\det\begin{bmatrix}
    \textbf B & -\textbf A\\
    -\textbf A & \textbf B
    \end{bmatrix}}}\exp\left(i\textbf d^T\textbf G\textbf d-\frac12i\begin{bmatrix}
    \textbf d^T\textbf{GS}+\textbf W^T & \textbf d^T\textbf{GS}-\textbf W^T
    \end{bmatrix}\begin{bmatrix}
    \textbf B & -\textbf A\\
    -\textbf A & \textbf B
    \end{bmatrix}^{-1}\begin{bmatrix}
    \textbf S^T\textbf G^T\textbf d+\textbf W \\ \textbf S^T\textbf G^T\textbf d-\textbf W
    \end{bmatrix}\right)
\end{equation}

First, the determinant in the square root can be expressed as $\det(\textbf B)\det(\textbf B-\textbf A\textbf B^{-1}\textbf A)$. Second, we can explicitly write out the inverse matrix in the exponential as follows:
\begin{equation}\label{explicit_inv}
    \begin{bmatrix}
        \textbf B & -\textbf A\\
        -\textbf A & \textbf B
    \end{bmatrix}^{-1}=\begin{bmatrix}
        (\textbf B-\textbf A\textbf B^{-1}\textbf A)^{-1} & (\textbf B-\textbf A\textbf B^{-1}\textbf A)^{-1}\textbf A\textbf B^{-1}\\
        (\textbf B-\textbf A\textbf B^{-1}\textbf A)^{-1}\textbf A\textbf B^{-1} & (\textbf B-\textbf A\textbf B^{-1}\textbf A)^{-1}
    \end{bmatrix}
\end{equation}

Using this identity, we can multiply out the matrix product in the exponential and obtain the following reduced form of Eq. \ref{intermediate_form}:
\begin{equation}
    \sqrt{\frac{\det(\bm a_e)\det(\bm a_g)}{\det(\textbf B)\det(\textbf B-\textbf{AB}^{-1}\textbf A)}}\exp\left(i\left[\textbf d^T\textbf G\textbf d-\textbf d^T\textbf{GS}(\textbf B-\textbf A)^{-1}\textbf S^T\textbf G^T\textbf d-\textbf W^T(\textbf B+\textbf A)^{-1}\textbf W\right]\right)
\end{equation}

Finally, for the sake of further simplification, we note that
\begin{align}
    &\textbf E+\textbf S^T\textbf{GS}=\textbf B-\textbf A\\
    \Rightarrow \quad&\textbf S^T\textbf G=(\textbf B-\textbf A)\textbf S^{\text{T}}-\textbf{ES}^T\\
    \Rightarrow \quad&\textbf d^T\textbf{GS}(\textbf B-\textbf A)^{-1}\textbf S^T\textbf G^T\textbf d=\textbf d^T\textbf G\textbf d-\textbf d^T\textbf{GS}(\textbf B-\textbf A)^{-1}\textbf{ES}^T\textbf d
\end{align}

Finally, the correlation function evaluates to
\begin{equation}\label{eq_corr_fn_derived}
    \sqrt{\frac{\det(\bm a_e)\det(\bm a_g)}{\det(\textbf B)\det(\textbf B-\textbf{AB}^{-1}\textbf A)}}\exp\left(i\left[\textbf d^T\textbf{GS}(\textbf B-\textbf A)^{-1}\textbf{ES}^T\textbf d-\textbf W^T(\textbf B+\textbf A)^{-1}\textbf W\right]\right)
\end{equation}

A reader familiar with the standard spin-boson model may wonder why we have not utilized the polaron transformation\cite{nitzanbook}, which is standard procedure in the derivation of Marcus rates. In the method outlined above, a polaron transformation will lead to exactly the same result obtained in Eq. \ref{eq_corr_fn_derived}. Although the polaron transformation would modify the interstate coupling operator, it would analogously modify the position basis used in the expansion of Eq. \ref{position_basis_expansion}.

\subsection{Derivation of Eq. \ref{neq_corr}}\label{app_neq}
The correlation function we now need to compute is
\begin{equation}
    C(t',t'') = \text{Tr}\left[e^{-\beta\hat H_B^{(g)}}e^{i\hat H_B^{(e)}t'}e^{-i\textbf W^T\hat{\textbf x}}e^{-i\hat H_B^{(g)}(t'-t'')}e^{i\textbf W^T \hat{\textbf x}}e^{-i\hat H_B^{(e)}t''}\right]
\end{equation}

Here, we must do 8 integrations over space instead of 4, with 4 excited state coordinate sets $\textbf x$, $\textbf y$, $\textbf z$, $\textbf w$ and 4 ground state coordinate sets $\textbf x'$, $\textbf y'$, $\textbf z'$, $\textbf w'$:
\begin{align}
    &\int_{\mathbb{R}^n}dx^n\,\int_{\mathbb{R}^n}dy^n\,\int_{\mathbb{R}^n}dz^n\,\int_{\mathbb{R}^n}dw^n\,\int_{\mathbb{R}^n}dx'^n\,\int_{\mathbb{R}^n}dy'^n\,\int_{\mathbb{R}^n}dz'^n\,\int_{\mathbb{R}^n}dw'^n\,\\
    &\braket*{\textbf x}{\textbf x'}\Braket*{\textbf x'}{e^{-\beta\hat H_B^{(g)}}}{\textbf y'}\braket*{\textbf y'}{\textbf y}
    \Braket*{\textbf y}{e^{i\hat H_B^{(e)}t'}e^{-i\textbf W^T\hat{\textbf x}}}{\textbf z}
    \braket*{\textbf z}{\textbf z'}
    \Braket*{\textbf z'}{e^{-i\hat H_B^{(g)}(t'-t'')}}{\textbf w'}
    \braket{\textbf w'}{\textbf w}    
    \Braket*{\textbf w}{e^{i\textbf W^T \hat{\textbf x}}e^{-i\hat H_B^{(e)}t''}}{\textbf x}\\ 
    =&\int_{\mathbb{R}^n}dx^n\,\int_{\mathbb{R}^n}dy^n\,\int_{\mathbb{R}^n}dz^n\,\int_{\mathbb{R}^n}dw^n\\
    &\Braket*{\textbf{Sx}+\textbf d}{e^{-\beta\hat H_B^{(g)}}}{\textbf{Sy}+\textbf d}
    \Braket*{\textbf y}{e^{i\hat H_B^{(e)}t'}}{\textbf z}
    \Braket*{\textbf{Sz}+\textbf d}{e^{-i\hat H_B^{(g)}(t'-t'')}}{\textbf{Sw}+\textbf d}    
    \Braket*{\textbf w}{e^{-i\hat H_B^{(e)}t''}}{\textbf x}   e^{i\textbf W^T {\textbf w}}e^{-i\textbf W^T {\textbf z}}
\end{align}

Now we look at the individual terms:
\begin{align}
    \Braket*{\textbf{Sx}+\textbf d}{e^{-\beta\hat H_B^{(g)}}}{\textbf{Sy}+\textbf d}&=\sqrt{\frac{\det \bm a_g(-i\beta)}{(2\pi i)^N}}\times\\
    &e^{i\left[\frac12\left((\textbf S\bm x+\textbf d)^{T}\bm b_g(-i\beta)(\textbf S\bm x+\textbf d)+(\textbf S\bm y+\textbf d)^{T}\bm b_g(-i\beta)(\textbf S\bm y+\textbf d)\right)-(\textbf S\bm x+\textbf d)^{T}\bm a_g(-i\beta)(\textbf S\bm y+\textbf d)\right]}\\
    \Braket*{\textbf y}{e^{i\hat H_B^{(e)}t'}}{\textbf z}&=\sqrt{\frac{\det \bm a_e(-t')}{(2\pi i)^N}}e^{i\left[\frac12\left(\bm y^{T}\bm b_e(-t')\bm y+\bm z^{T}\bm b_e(-t')\bm z\right)-\bm y^{T}\bm a_e(-t')\bm z\right]}\\
    \Braket*{\textbf{Sz}+\bm\gamma}{e^{-i\hat H_B^{(g)}(t'-t'')}}{\textbf{Sw}+\bm\gamma}&=\sqrt{\frac{\det \bm a_g(t'-t'')}{(2\pi i)^N}}\times\\
    &e^{i\left[\frac12\left((\textbf{S}\bm z+\textbf d)^{T}\bm b_g(t'-t'')(\textbf{S}\bm z+\textbf d)+(\textbf{S}\bm w+\textbf d)^{T}\bm b_g(t'-t'')(\textbf{S}\bm w+\textbf d)\right)-(\textbf{S}\bm z+\textbf d)^{T}\bm a_g(t'-t'')(\textbf{S}\bm w+\textbf d)\right]}\\
    \Braket*{\textbf w}{e^{i\hat H_B^{(e)}t'}}{\textbf x}&=\sqrt{\frac{\det \bm a_e(t'')}{(2\pi i)^N}}e^{i\left[\frac12\left(\bm w^{T}\bm b_e(t'')\bm w+\bm x^{T}\bm b_e(t'')\bm x\right)-\bm w^{T}\bm a_e(t'')\bm x\right]}
\end{align}

Collecting bilinear and linear terms, we can write the integrand (sans prefactor) as
\begin{align}
    &\exp\left(i\left[\frac12
    \begin{bmatrix}
    \bm x^{T}&\bm y^{T}&\bm z^{T}&\bm w^{T}
    \end{bmatrix}
    \SSigma
    \begin{bmatrix}
    \bm x\\
    \bm y\\
    \bm z\\
    \bm w
    \end{bmatrix}
    +\begin{bmatrix}
    \bm x^{T}&\bm y^{T}&\bm z^{T}&\bm w^{T}
    \end{bmatrix}\begin{bmatrix}
    \textbf S^T\textbf{G}_\beta\textbf d\\
    \textbf S^T\textbf{G}_\beta\textbf d\\
    \textbf S^{T}\textbf G\textbf d-\textbf W\\
    \textbf S^{T}\textbf G\textbf d+\textbf W
    \end{bmatrix}
    +\textbf d^{T}\left(\textbf{G}_\beta+\textbf G\right)\textbf d
    \right]\right)
\end{align}
where
\begin{align}
    &\SSigma = \begin{bmatrix}
    \textbf S^T\bm b_g(-i\beta)\textbf S+\bm b_e(t'')&-\textbf S^T\bm a_g(-i\beta)\textbf S&0&-\bm a_e(t'')\\
    -\textbf S^T\bm a_g(-i\beta)\textbf S&\textbf S^T\bm b_g(-i\beta)\textbf S+\bm b_e(-t')&-\bm a_e(-t')&0\\
    0&-\bm a_e(-t')&\bm b_e(-t')+\textbf S^{T}\bm b_g(t'-t'')\textbf S&-\textbf S^{T}\bm a_g(t'-t'')\textbf S\\
    -\bm a_e(t'')&0&-\textbf S^{T}\bm a_g(t'-t'')\textbf S &\bm b_e(t'')+\textbf S^{T}\bm b_g(t'-t'')\textbf S
    \end{bmatrix}\\
    &\textbf{G} = \bm b_g(t'-t'')-\bm a_g(t'-t'')\\
    &\textbf{G}_\beta = \bm b_g(-i\beta)-\bm a_g(-i\beta)
\end{align}

As before, we switch to a single integral over a $4n$-dimensional space:
\begin{align}
    \int_{\mathbb{R}^{4n}}d\alpha^{4n}\,\exp\left(-\frac12
    \bm \alpha^{T}
    \left(-i\SSigma\right)
    \bm\alpha
    +i\,\bm\alpha^{T}
    \begin{bmatrix}
    \textbf S^T\textbf{G}_\beta\textbf d\\
    \textbf S^T\textbf{G}_\beta\textbf d\\
    \textbf S^{T}\textbf G\textbf d-\textbf W\\
    \textbf S^{T}\textbf G\textbf d+\textbf W
    \end{bmatrix}
    +i\,\textbf d^{T}\left(\textbf{G}_\beta+\textbf G\right)\textbf d
    \right)
\end{align}

Finally, the integral (with prefactors) evaluates to
\begin{align}\label{final_corrfn}
    \sqrt{\frac{\det\left[\bm a_g(-i\beta)\bm a_e(-t')\bm a_g(t'-t'')\bm a_e(t'')\right]}{\det\Sigma}}\exp\left(i\,\textbf d^{T}\left(\textbf{G}_\beta+\textbf G\right)\textbf d-\frac12i\begin{bmatrix}
    \textbf S^T\textbf{G}_\beta\textbf d\\
    \textbf S^T\textbf{G}_\beta\textbf d\\
    \textbf S^{T}\textbf G\textbf d-\textbf W\\
    \textbf S^{T}\textbf G\textbf d+\textbf W
    \end{bmatrix}^{T}\SSigma^{-1}\begin{bmatrix}
    \textbf S^T\textbf{G}_\beta\textbf d\\
    \textbf S^T\textbf{G}_\beta\textbf d\\
    \textbf S^{T}\textbf G\textbf d-\textbf W\\
    \textbf S^{T}\textbf G\textbf d+\textbf W
    \end{bmatrix}\right)
\end{align}

\subsection{A Note on Numerical Implementation}\label{implementation}
    In the Hamiltonian defined in Eq. \ref{langevin_H}, the number of bath coordinates is theoretically infinite, resulting in a bath with a continuous spectrum. Furthermore, the frequencies present in that spectrum are unbounded (i.e. the bath contains modes of arbitrarily large frequency). In order to perform a Fermi's Golden Rule calculation, we must truncate and discretize this bath. The procedure used for our simulations is as follows: First, we choose a cutoff frequency $\omega_C$. Then, we choose $M$ frequencies evenly spaced in the interval $(0,\omega_C)$ and calculate the density of modes as $\omega_C/M$. Finally, these parameters can be used to calculate the harmonic matrices required in Eq. \ref{langevin_H_transformed}. Note that one need not use a uniform patterning of frequency space, and a nonuniform discretization may well provide faster convergence of computed dynamics as the number of modes is increased. However, a uniform discretization proved to be adequate for our needs. In this scheme, good convergence can be achieved with as few as 40 bath modes.

\printbibliography
\end{document}